\documentclass[a4paper,11pt]{article}
\pdfoutput=1
\usepackage{jheppub}
\usepackage[T1]{fontenc}
\usepackage{tensor}
\usepackage{bbm}
\usepackage{parskip}
\usepackage{physics}
\newcommand{\mn}{\mu\nu}

\newcommand{\inv}[1]{\frac{1}{#1}}

\renewcommand{\cal}[1]{\mathcal{#1}}
\newcommand{\brac}[1]{\left(#1\right)}
\newcommand{\hpart}{\hat{\partial}}
\newcommand{\bbm}[1]{\mathbbm{#1}}
\newcommand{\bb}[1]{\mathbb{#1}}
\newcommand{\vac}[1]{\bra{\Omega} #1 \ket{\Omega} }
\renewcommand{\frak}[1]{\mathfrak{#1}}
\newcommand{\polylog}[2]{\mathrm{Li}_{#1}(#2)}
\DeclareMathOperator{\Exp}{Exp}

\allowdisplaybreaks

\usepackage{hyperref}
\title{Fermions with \boldmath $SU(1,n)$ Spacetime Symmetry}

\author{Joseph Smith}
\affiliation{Department of Mathematics, King's College London,\\The Strand, London WC2R 2LS, U.K.}
\emailAdd{joseph.m.smith@kcl.ac.uk}

\abstract{We construct theories of free fermions in $(2n-1)$-dimensions with $SU(1,n)$ spacetime symmetry from the null reduction of fermions on a $2n$-dimensional $\Omega$-deformed Minkowski background for $n=2$ and $n=3$. These play a role in the 5d $SU(1,3)$-invariant theories that are conjectured to offer a full description of certain 6d superconformal field theories. We find the $(2n-1)$-dimensional manifestation of the supersymmetry of a free $2n$-dimensional boson-fermion system, which we use to fix the fermion two-point functions. It is then shown that the full $2n$-dimensional two-point function can be recovered through resummation. Limits of the theories are considered, and it is observed that both Galilean and Carrollian field theories appear in different regimes.
We confirm that the correlation functions obey the $SU(1,n)$ Ward identities and the representations of the fermions under this group are discussed.}

\begin{document} 
\maketitle
\flushbottom

\section{Introduction}
\label{sec: intro}

An interesting theme in modern research has been the study of field theories without Lorentz symmetry. While these have long been considered in the context of condensed matter physics, there have been numerous recent advances in their application to problems in high energy physics. These include non-relativistic limits of string theory and supergravity \cite{Gomis:2000bd, Gomis:2005pg,  Bergshoeff:2023ogz,Oling:2022fft,Harmark:2019upf}, 
the identification of non-relativistic limits of $\cal{N}=4$ super Yang-Mills with quantum mechanical models \cite{Harmark:2014mpa, Baiguera:2020jgy, Baiguera:2022pll}, and the role of Carrollian field theories in flat-space holography \cite{Donnay:2022aba, Donnay:2022wvx, Bagchi:2022emh, Duval:2014lpa, Duval:2014uva}, as well as many other topics.

Though non-Lorentzian field theories have an existence outside of Lorentzian theories, the two are often connected. For instance, non-Lorentzian field theories can be constructed from a Lorentzian one through null reduction. One of the lightcone directions in $D$-dimensional Minkowski spacetime is periodically identified and a Kaluza-Klein reduction on this circle is performed in a procedure known as Discrete Lightcone Quantization (DLCQ). The theory can then be truncated at a given level, leaving us with a theory in $D-1$ dimensions with a non-Lorentzian spacetime symmetry. If the theory we started with was conformally invariant, the reduced theory has Schr\"odinger symmetry \cite{Son:2008ye}.

The $D-1$ dimensional theory we obtain at a given level is well-defined, but the question of whether the full $D$-dimensional theory can be recovered from the Kaluza-Klein tower is subtle \cite{Yamawaki:1998cy} and we may wonder if we can find a  similar procedure that resolves these issues. In \cite{Lambert:2020zdc}, a diffeomorphism of Minkowski spacetime in $D=2n$ dimensions was introduced that, after a conformal compactification, maps to an $\Omega$-deformed \cite{Nekrasov:2002qd} spacetime in which one of the null directions takes values in a finite interval. Reducing a theory on this direction leaves us with a Kaluza-Klein tower of theories with $SU(1,n)$ spacetime symmetry \cite{Lambert:2021nol}. Since the deformation is induced using a combination of coordinate and Weyl transformations, we can map quantities computed in the deformed theory to the original Lorentzian theory (and vice-versa). This is unlike the DLCQ approach, where the procedure to recover the original theory (decompactification of the null circle) is not invertible.
Keeping the full tower of $SU(1,n)$-invariant $(2n-1)$-dimensional field theories should allow us, in principle, to be able to compute quantities in the original theory that are inaccessible in other approaches. Even outside of this connection to Lorentzian field theories, $SU(1,n)$ theories are a novel class of QFTs that are worthy of further study, particularly with respect to their quantum properties. For instance, it was shown in \cite{Lipstein:2022kre} that the 5d superconformal group with bosonic subgroup $SU(1,3)\times U(1) \times SO(5)$ admits an infinite-dimensional Yangian extension. It is therefore possible that there exist integrable limits of theories with the superconformal symmetry. 

A noteworthy application of $SU(1,n)$ theories has been to the study of 6d superconformal field theories (SCFTs). As such theories are not thought to admit conventional 6d Lagrangian descriptions, most of what is known about them can only be found from string constructions (see \cite{Heckman:2018jxk} for a review). In \cite{Lambert:2019jwi, Lambert:2020jjm} 5d Lagrangian theories with $SU(1,3)$ spacetime symmetry were introduced and conjectured to provide a complete description of certain 6d SCFTs. While an exciting prospect, there are still many non-trivial properties these theories must possess. For example, the Lagrangians are only invariant under half the supercharges\footnote{Note that we are referring to non-conformal supercharges here.} of the 6d theories. The proposal requires that the remaining half appear non-perturbatively, with the supercharges taking us between levels of the Kaluza-Klein tower given by sectors of different instanton number. This is difficult to verify and it would be useful to study a toy model in which we can explicitly show how all the $2n$-dimensional supercharges act in the reduced theory. This can be provided by the reduction of a $2n$-dimensional supersymmetric theory where we have access to the full tower of fields at the Lagrangian level. The simplest example we can consider is a free theory comprising a complex scalar and a Weyl spinor. Previous work \cite{Lambert:2020zdc, Lambert:2021nol} has focused on the properties of scalar fields with $SU(1,n)$ spacetime symmetry and relatively little is known about fermions in these backgrounds. The construction and analysis of such theories is therefore of interest, and will be the focus of this work.

The paper is organised as follows. In section \ref{sec: action} we review the coordinate transformation that induces the $\Omega$-deformation and find the action of a $2n$-dimensional free Dirac fermion on this background, which we then reduce along the null direction to obtain a tower of $(2n-1)$-dimensional theories. In section \ref{sec: susy and corr} we examine the manifestation of $2n$-dimensional supersymmetry in the $(2n-1)$-dimensional theories; we then use this to determine the fermionic two-point functions using the scalar two-point functions determined in \cite{Lambert:2021nol}. The tower of $(2n-1)$-dimensional correlation functions are resummed, reproducing the two-point function of the $2n$-dimensional theory. In section \ref{sec: limits} we examine a limit of the $(2n-1)$-dimensional theory that reduces it to the DLCQ of a free massless fermion. We find the two-point functions of the $(2n-1)$-dimensional theory in this limit and show that the original $2n$-dimensional two-point function can still be recovered. We also comment on a related limit that gives a theory with Carroll symmetry. The reduction of a 2d fermion is considered in section \ref{sec: 2d ferm reduction}. In section \ref{sec: WI} we show that the $(2n-1)$-dimensional two-point functions satisfy the Ward identities associated with $\frak{su}(1,n)$ spacetime symmetry. Finally, in section \ref{sec: conclusion} we discuss our results and prospects for future work. Various additional material is collected in the appendices. In appendix \ref{sec: conv} we outline our conventions for fermions and gamma matrices. In appendix \ref{sec: gen} we provide a derivation of the generator of the local Lorentz transformation associated with our coordinate transformation. In appendix \ref{sec: scalar coefficients} the coefficients of the two-point functions for $SU(1,n)$-invariant scalar fields are computed and shown to satisfy an infinite tower of Green's function-like equations that descend from the $2n$-dimensional Green's function equation.

\section{Reduction of the Fermionic Action}
\label{sec: action}

Following the philosophy of \cite{Lambert:2021nol}, we will construct fermions with $SU(1,n)$ spacetime symmetry from the reduction of free fermions in $2n$-dimensional Minkowski spacetime along a null interval\footnote{The fermion action for generic $n$ was presented in passing in \cite{Lambert:2021nol}, but key details about the transformation of the fields that will be necessary to reconstruct the $2n$-dimensional theory were not emphasised.}. This leads to an infinite tower of decoupled theories in $(2n-1)$-dimensions. Depending on one's motivation we can either truncate to a finite set of fields that possess $SU(1,n)$ spacetime symmetry or we can keep the whole Kaluza-Klein tower, allowing us to reconstruct the original $2n$-dimensional theory. We will take the latter approach, which will impose extra constraints on the moding of our fields: such constraints can be dropped if one is only interested in intrinsically defined $(2n-1)$-dimensional theories. Our goal will ultimately be to show that the Kaluza-Klein tower of $(2n-1)$-dimensional fermions is equivalent to the original $2n$-dimensional theory and that no information is lost in this procedure.

\subsection{Coordinate Transformation}
\label{subsec: coord trans}

Let us review the coordinate transformation introduced in \cite{Lambert:2021nol}. We start with Minkowski spacetime in $D =2n$ dimensions and split our coordinates into lightcone directions $(\hat{x}^+, \hat{x}^-)$ and transverse directions $\hat{x}^i$. Though we will keep our dimension general at this stage, we will be interested in the cases $n=2$ and $n=3$ as these are the only (even) dimensions in which we have both transverse coordinates and known interacting CFTs; for this reason, when explicit expressions are required we will only provide them for these cases.

We work with the coordinate transformation
\begin{subequations} \label{eq: coord transformation}
\begin{align}
    \hat{x}^+ &= 2R \tan\brac{\frac{x^+}{2R}} \ , \\
    \hat{x}^- &= x^- + \frac{x^2}{4R} \tan\brac{\frac{x^+}{2R}} \ , \\
    \hat{x}^i &= x^i - R \Omega_{ij} x^j \tan\brac{\frac{x^+}{2R}} \ ,
\end{align}
\end{subequations}
defined in terms of a lengthscale $R$ and an antisymmetric matrix $\Omega_{ij}$ satisfying
\begin{equation}
    \Omega_{ij} \Omega_{ik} = \inv{R^2} \delta_{jk}
\end{equation}
that acts on the transverse coordinates. We will use the notation
\begin{equation}
    x^i x^i \equiv x^2
\end{equation}
throughout when this causes no confusion. For convenience, we will choose $\Omega_{ij}$ to be
\begin{equation}
    \Omega^{(4d)}_{ij} = \inv{R} \begin{pmatrix}
        0 & 1 \\
        -1 & 0
    \end{pmatrix}
\end{equation}
in 4d and
\begin{equation}
    \Omega^{(6d)}_{ij} = \inv{R} \begin{pmatrix}
        0 & 1 & 0 & 0 \\
        -1 & 0 & 0 & 0 \\
        0 & 0 & 0 & 1 \\
        0 & 0 & -1 & 0
    \end{pmatrix}
\end{equation}
in 6d\footnote{We note that with this choice $\Omega^{(6d)}$ is self-dual, in contrast to the anti-self-dual choice made in previous work.} when performing explicit computations.

With this choice the coordinate $x^+$ takes values in the interval $(-\pi R, \pi R)$. After a Weyl transformation of the metric we can compactify the interval through the inclusion of its endpoints, extending the range of $x^+$ to $[-\pi R, \pi R]$. The transformation of the metric takes the form
\begin{align} \nonumber
    \hat{g}_{\mn} d\hat{x}^{\mu} d\hat{x}^{\nu} &=  -2 d\hat{x}^+ d\hat{x}^- + d\hat{x}^i d\hat{x}^i \\ \nonumber
    &= \Delta^2 \brac{-2 dx^+ \brac{dx^- + \inv{2}\Omega_{ij} x^j dx^i} + dx^i dx^i } \\ \label{eq: metric}
    &= \Delta^2 g_{\mn} dx^{\mu} dx^{\nu} \ ,
\end{align}
with
\begin{equation}
    \Delta = \sec \brac{\frac{x^+}{2R}} \ .
\end{equation}
From our perspective the choice of the scale $R$ is arbitrary; however, we note that in the holographic realisation of this geometry \cite{Lambert:2019jwi} $R$ is the $AdS_{2n+1}$ radius.

Our goal is to work with spinor fields on the $\Omega$-deformed geometry defined above, so we need to define vielbeins $\hat{e}$ and $e$ for the metrics $\hat{g}$ and $g$. If we apply a combination of the coordinate transformation \eqref{eq: coord transformation} and a Weyl transformation by $\Delta$ to $\hat{e}$ we will in general not obtain $e$, but a vielbein related to it by a local Lorentz transformation $\Lambda\in SO(1,2n-1)$ given by
\begin{equation} \label{eq: vielbein consistency}
    \tensor{\Lambda}{^a_b} = \Delta \frac{\partial x^{\mu}}{\partial \hat{x}^{\nu}} e^{a}_{\mu} \hat{e}^{\nu}_b \ .
\end{equation}
Unlike in the bosonic case, $\Lambda$ will enter into the transformation of a spinor field and will be crucial in relating correlation functions of the theory defined on the $\Omega$-deformed background to those of the original Minkowski theory.

A convenient choice for the vielbeins is
\begin{subequations}
\begin{align}
    \hat{e}^0 &= \inv{\sqrt{2}} \brac{d\hat{x}^+ + d\hat{x}^-} \ , \\
    \hat{e}^1 &= \inv{\sqrt{2}} \brac{d\hat{x}^+ - d\hat{x}^-} \ , \\
    \hat{e}^i &= d\hat{x}^i \ ,
\end{align}
\end{subequations}
for the metric $\hat{g}$, and
\begin{subequations} \label{eq: e veilbein}
\begin{align}
    e^0 &= \inv{\sqrt{2}} \brac{dx^+ + dx^- + \inv{2} \Omega_{ij}x^j dx^i} \ , \\
    e^1 &= \inv{\sqrt{2}} \brac{dx^+ - dx^- - \inv{2} \Omega_{ij}x^j dx^i} \ , \\
    e^i &= dx^i \ ,
\end{align}
\end{subequations}
for the metric $g$. With these in hand, a brief calculation shows the components of $\Lambda$ are
\begin{subequations} \label{eq: lambda}
\begin{align}
    \tensor{\Lambda}{^0_0} &= \inv{2} \brac{\cos\brac{\frac{x^+}{2R}} + \sec\brac{\frac{x^+}{2R}} \left[1 + \frac{x^2}{8R^2} \right]} \ , \\
    \tensor{\Lambda}{^1_0} &= \inv{2} \brac{\cos\brac{\frac{x^+}{2R}} - \sec\brac{\frac{x^+}{2R}} \left[1 + \frac{x^2}{8R^2} \right]} \ , \\
    \tensor{\Lambda}{^0_1} &= \inv{2} \brac{\cos\brac{\frac{x^+}{2R}} - \sec\brac{\frac{x^+}{2R}} \left[1 - \frac{x^2}{8R^2} \right]} \ , \\
    \tensor{\Lambda}{^1_1} &= \inv{2} \brac{\cos\brac{\frac{x^+}{2R}} + \sec\brac{\frac{x^+}{2R}} \left[1 - \frac{x^2}{8R^2} \right]} \ , \\
    \tensor{\Lambda}{^i_0} &= \tensor{\Lambda}{^i_1} = \inv{2\sqrt{2}} \brac{ \cos\brac{\frac{x^+}{2R}} \Omega_{ij} x^j  -  \sin\brac{\frac{x^+}{2R}} \frac{x^i}{R} } \ , \\
    \tensor{\Lambda}{^0_j} &= - \tensor{\Lambda}{^1_j} = \inv{2\sqrt{2}} \sec\brac{\frac{x^+}{2R}} \, \Omega_{jk}x^k \ , \\
    \tensor{\Lambda}{^i_j} &= \cos\brac{\frac{x^+}{2R}} \, \delta_{ij} + \sin{\brac{\frac{x^+}{2R}}} \, R \Omega_{ij} \ .
\end{align}
\end{subequations}

We are interested in transforming spinor fields from Minkowski spacetime to our new coordinate system. In doing so we will require a spinor representation of the local Lorentz transformation $\Lambda$. The simplest spinor representation to compute is the Dirac representation, where $\Lambda$ is represented by
\begin{equation}
    S[\Lambda] = \Exp \brac{\inv{4} \lambda^{ab} \gamma_{ab}} \ .
\end{equation}
Calculating this requires knowledge of the Lie algebra element $\lambda \in \frak{so}(1,2n-1)$ that generates $\Lambda$, which in our conventions is given by
\begin{equation}
    \Lambda = \Exp\brac{\lambda} \ .
\end{equation}

While it is not immediately obvious what form this should take, a derivation based on the accidental isomorphism between $Spin(1,3)$ and $SL(2,\bb{C})$ (given in appendix \ref{sec: gen}) for the $n=2$ case or an inspired guess shows that the correct element is
\begin{subequations}
\begin{align}
    \lambda^{01} &= \ln\cos\brac{\frac{x^+}{2R}} \ , \\
    \lambda^{ij} &= \inv{2} \Omega_{ij} x^+ \ ,
\end{align}
\begin{align} \nonumber
    \lambda^{0i} = - \lambda^{1i} = - \inv{2\sqrt{2}R \sin\brac{\frac{x^+}{2R}}}
    \Bigg[x^i \brac{\cos\brac{\frac{x^+}{2R}} \ln\cos\brac{\frac{x^+}{2R}} + \sin\brac{\frac{x^+}{2R}} \frac{x^+}{2R} }& \\ 
    + R \Omega_{ij}x^j \brac{\sin\brac{\frac{x^+}{2R}} \ln\cos\brac{\frac{x^+}{2R}} - \cos\brac{\frac{x^+}{2R}} \frac{x^+}{2R}} \Bigg]&
     \ .
\end{align}
\end{subequations}
It is then a straightforward task to compute the Dirac spinor representation of $\Lambda$; using the gamma matrices outlined in appendix \ref{sec: conv} we find
\begin{subequations} \label{eq: S for 4d}
\begin{align}
    S^{(4d)} &= \begin{pmatrix}
        S^{(4d)}_+ & 0 \\
        0 & S^{(4d)}_-
    \end{pmatrix} \ , \\
    S^{(4d)}_+ &= \frac{e^{-\frac{i x^+}{4R}}}{4\sqrt{\cos\brac{ \frac{x^+}{2R} }}} \begin{pmatrix}
        3 + e^{\frac{i x^+}{R}} - \frac{u}{\sqrt{2}R} & -1 + e^{\frac{i x^+}{R}} - \frac{u}{\sqrt{2}R} \\
        -1 + e^{\frac{i x^+}{R}} + \frac{u}{\sqrt{2}R} & 3 + e^{\frac{i x^+}{R}} + \frac{u}{\sqrt{2}R}
    \end{pmatrix} \ , \\
    S^{(4d)}_- &= \frac{e^{\frac{i x^+}{4R}}}{4\sqrt{\cos\brac{ \frac{x^+}{2R} }}} \begin{pmatrix}
        3 + e^{-\frac{i x^+}{R}} + \frac{\bar{u}}{\sqrt{2}R} & 1 - e^{-\frac{i x^+}{R}} - \frac{\bar{u}}{\sqrt{2}R} \\
        1 - e^{-\frac{i x^+}{R}} + \frac{\bar{u}}{\sqrt{2}R} & 3 + e^{-\frac{i x^+}{R}} - \frac{\bar{u}}{\sqrt{2}R}
    \end{pmatrix} \ ,
\end{align}
\end{subequations}
for the representation in 4d, and
\begin{subequations} \label{eq: S for 6d}
\begin{align}
    S^{(6d)}[\Lambda] &= \begin{pmatrix}
        S^{(6d)}_+ & 0 \\
        0 & S^{(6d)}_-
    \end{pmatrix} \ , \\ \nonumber
    S^{(6d)}_+ &= \inv{4 \sqrt{\cos\brac{ \frac{x^+}{2R} }}} \begin{pmatrix}
        3 + e^{\frac{i x^+}{R}} & -1 + e^{\frac{i x^+}{R}}  &  0 & 0 \\
        -1 + e^{\frac{i x^+}{R}} &  3 + e^{\frac{i x^+}{R}} & 0 & 0 \\
        0 & 0 &  3 + e^{-\frac{i x^+}{R}} &  1 - e^{-\frac{i x^+}{R}} \\
        0 & 0 & 1 - e^{-\frac{i x^+}{R}} & 3 + e^{-\frac{i x^+}{R}} \\
    \end{pmatrix} \\
    &\hspace{0.5cm}+ \inv{4\sqrt{2}R \sqrt{\cos\brac{ \frac{x^+}{2R} }}} \begin{pmatrix}
        -u & -u & v & -v \\
        u & u & -v & v \\
        \bar{v} & \bar{v} & \bar{u} & - \bar{u} \\
        \bar{v} & \bar{v} & \bar{u} & - \bar{u}
    \end{pmatrix}
    \ , \\ \nonumber
    S_-^{(6d)} &= \frac{e^{-\frac{i x^+}{2R}}}{4\sqrt{\cos\brac{ \frac{x^+}{2R} }}} 
    \begin{pmatrix}
        3 + e^{\frac{i x^+}{R}} - \frac{u}{\sqrt{2}R} & -1 + e^{\frac{i x^+}{R}} - \frac{u}{\sqrt{2}R} &  - \frac{\bar{v}}{\sqrt{2}R} &  \frac{\bar{v}}{\sqrt{2}R} \\
        -1 + e^{\frac{i x^+}{R}} + \frac{u}{\sqrt{2}R} &  3 + e^{\frac{i x^+}{R}} + \frac{u}{\sqrt{2}R} & \frac{\bar{v}}{\sqrt{2}R} & - \frac{\bar{v}}{\sqrt{2}R} \\
        0 & 0 & 0 & 0 \\
        0 & 0 & 0 & 0
    \end{pmatrix} \\ 
    &\hspace{0.5cm} + \frac{e^{\frac{i x^+}{2R}}}{4\sqrt{\cos\brac{ \frac{x^+}{2R} }}} 
    \begin{pmatrix}
        0 & 0 & 0 & 0 \\
        0 & 0 & 0 & 0 \\
        -\frac{v}{\sqrt{2}R} & -\frac{v}{\sqrt{2}R} &  3 + e^{-\frac{i x^+}{R}} + \frac{\bar{u}}{\sqrt{2}R} &  1 - e^{-\frac{i x^+}{R}} - \frac{\bar{u}}{\sqrt{2}R} \\
        -\frac{v}{\sqrt{2}R} & -\frac{v}{\sqrt{2}R} & 1 - e^{-\frac{i x^+}{R}} + \frac{\bar{u}}{\sqrt{2}R} & 3 + e^{-\frac{i x^+}{R}} - \frac{\bar{u}}{\sqrt{2}R}\\
    \end{pmatrix} \ ,
\end{align}
\end{subequations}
for the 6d case. These are written in terms of the complex coordinates
\begin{subequations}
\begin{gather}
    u = x^2 + i x^3 \ , \\
    v = x^5 + i x^4 \ .
\end{gather}
\end{subequations}
Our choice of gamma matrices makes the chiral decomposition of the Dirac representation manifest, which will be convenient when working with Weyl spinors.

The last quantity we require is the spin connection $\tensor{\omega}{_a_b}$ of \eqref{eq: e veilbein}. A quick computation gives
\begin{subequations} \label{eq: spin conn}
\begin{gather}
    \omega_{0i} = \inv{2\sqrt{2}} \Omega_{ij} dx^j \ , \\
    \omega_{1i} = \inv{2\sqrt{2}} \Omega_{ij} dx^j \ , \\
    \omega_{ij} = - \inv{2} \Omega_{ij} dx^+ \ .
\end{gather}
\end{subequations}

\subsection{The Action for General Dimensions} \label{sec: action 2}

With the geometric quantities calculated above in hand, we are ready to find the action of a fermion on the $\Omega$-deformed background. The action for a free massless Dirac spinor on a curved background with metric $\hat{g}$ is
\begin{equation}
    S = - \int d^D \hat{x} \, \hat{e} \hat{\bar{\psi}} \gamma^a \hat{e}_a^{\mu} \hat{\nabla}_{\mu} \hat{\psi} \ .
\end{equation}
We are interested in performing a coordinate transformation of the form considered above, in which we write the transformed metric as
\begin{equation}
    \hat{g}_{\mn}(\hat{x}) d\hat{x}^{\mu} d\hat{x}^{\nu} = \Delta^2(x) g_{\mn}(x) dx^{\mu} dx^{\nu} \ .
\end{equation}
Specifying a vielbein $e$ for $g$, we can use the Lorentz transformation in eq. \eqref{eq: vielbein consistency} to define the spinor field transformation
\begin{equation} \label{eq: spinor transformation}
    \hat{\psi}(\hat{x}) = \Delta^{- \frac{D-1}{2}} S^{-1}[\Lambda] \psi(x) \ .
\end{equation}
It can be shown that this choice leaves the form of the action invariant \cite{Shapiro:2016pfm}, so we are left with
\begin{equation}
    S = - \int d^{D}x \, e \bar{\psi} \gamma^a e^{\mu}_a \nabla_{\mu} \psi \ .
\end{equation}

Let us now specialise to the transformation \eqref{eq: coord transformation} in $D=2n$ dimensions. We can substitute \eqref{eq: e veilbein} and \eqref{eq: spin conn} into the action to obtain the $\Omega$-deformed spinor action
\begin{equation}
    S = i \int d^{2n} x \, \psi^{\dag} \Bigg[ \sqrt{2} P_+ \brac{\partial_+ + \inv{8} \Omega_{ij} \gamma_{ij} } \psi + \sqrt{2} P_- \partial_- \psi + \gamma_{0i} D_i \psi 
    \Bigg] \ ,
\end{equation}
where we work with the projectors $P_{\pm} = \inv{2} \brac{\bbm{1} \pm \gamma_{01}}$ and have defined the differential operator
\begin{equation}
    D_i = \partial_i - \inv{2} \Omega_{ij} x^j \partial_- \ .
\end{equation}
 In this form the complete symmetry group of the action after null reduction will not be manifest: we can remedy this by introducing the projections of $\psi$ with respect to $P_{\pm}$, which we denote by
\begin{equation}
    P_+ \psi = \chi \; , \; P_- \psi = \lambda \ .
\end{equation}
The action is then
\begin{equation} \label{eq: action 1}
    S = i \int d^{2n} x \, \Bigg[ \sqrt{2} \lambda^{\dag} \partial_- \lambda + \chi^{\dag} \gamma_{0i} D_i \lambda + \lambda^{\dag} \gamma_{0i} D_i \chi + \sqrt{2} \chi^{\dag} \brac{\partial_+ + \inv{8} \Omega_{ij} \gamma_{ij} } \chi \Bigg] \ .
\end{equation}

We would like to reduce our theory on the null interval. To do this, we take the fields to have mode expansions of the form
\begin{subequations}
\begin{gather}
    \lambda(x^+,x^-,x) = \sum_{k\in \cal{S}} \lambda_k(x^-,x) e^{- \frac{i k x^+}{R}} \ , \\
    \chi(x^+,x^-,x) = \sum_{k\in \cal{S}} \chi_k(x^-,x) e^{- \frac{i k x^+}{R}} \ .
\end{gather}
\end{subequations}
The set of allowed mode numbers $\cal{S}$ is fixed by requiring that the field obtained from a coordinate transformation of the Minkowski spinor,
\begin{equation}
    \hat{\psi}(\hat{x}) = \psi'(x) \ ,
\end{equation}
can be written as the Fourier series
\begin{equation}
    \psi'(x^+,x^-,x) = \sum_{l\in\bb{Z}} \psi'_l(x^-,x) e^{- \frac{i l x^+}{R}} \ ,
\end{equation}
mirroring the analogous prescription for a scalar field \cite{Lambert:2021nol}. This amounts to demanding consistency of the equation
\begin{equation} \label{eq: consistency}
    \Delta^{-\frac{d-1}{2}} S^{-1}[\Lambda] \sum_{k\in \cal{S}} \psi_k(x^-,x) e^{- \frac{i k x^+}{R}} = \sum_{l\in\bb{Z}} \psi'_l(x^-,x) e^{- \frac{i l x^+}{R}} \ .
\end{equation}
As the Lorentz transformation is dependent on $x^+$ the details of this will depend on the spacetime dimension. Substituting the mode expansion into \eqref{eq: action 1} allows us to perform the integral over $x^+$, giving
\begin{equation} \label{eq: action 2}
    S = 2\pi i R \sum_{k\in\cal{S}}  \int d^{2n-1}x \Bigg[ \sqrt{2} \lambda_k^{\dag} \partial_- \lambda_k + \chi_k^{\dag} \gamma_{0i} D_i \lambda_k + \lambda_k^{\dag} \gamma_{0i} D_i \chi_k + \sqrt{2} \chi^{\dag} \brac{ \inv{8} \Omega_{ij} \gamma_{ij} - \frac{ik}{R} } \chi_k \Bigg] \ .
\end{equation}
This action was first constructed in \cite{Lambert:2021nol}. We note that if we were only interested in using the reduction to construct a fermion with $SU(1,n)$ spacetime symmetry (i.e. if we were to truncate to the pair $(\lambda_k, \chi_k)$) then we can relax the condition on $k$ and instead take it to be any real number.

\subsubsection{The Action for 5d}

Let us now specialise \eqref{eq: action 2} to specific dimensions, namely $D=6$ and $D=4$. While most of \eqref{eq: action 2} is independent of dimension, the term involving $\Omega_{ij} \gamma_{ij}$ is not. Expanding out the expression for $D=6$ allows us to rewrite this as
\begin{equation}
    \Omega_{ij} \gamma_{ij} = \frac{2}{R} \gamma_{23} \brac{\bbm{1} + \gamma_{01} \gamma_*} \ ,
\end{equation}
where our convention for $\gamma_*$ is given in appendix \ref{sec: conv}. Let us now impose the condition that our fields are Weyl spinors with negative chirality: from this we see that
\begin{equation}
    \Omega_{ij} \gamma_{ij} \chi = 0
\end{equation}
and the final term in the action vanishes, leaving the simplified 5d action
\begin{equation}
    S^{(5d)} = 2\pi i R \sum_{k\in\cal{S}^{(5d)}} \int d^5x \Bigg[ \sqrt{2} \lambda_k^{\dag} \partial_- \lambda_k + \chi_k^{\dag} \gamma_{0i} D_i \lambda_k + \lambda_k^{\dag} \gamma_{0i} D_i \chi_k - \frac{\sqrt{2} i k}{R} \chi^{\dag}_k \chi_k
    \Bigg] \ .
\end{equation}

Using the explicit form \eqref{eq: S for 6d} of the spinor representation, the consistency condition \eqref{eq: consistency} for $\cal{S}^{(5d)}$ can be written as
\begin{equation}
    \sum_{l \in \bb{Z}} \psi'_{l} e^{-\frac{i l x^+}{R}} = \brac{ e^{\frac{2i x^+}{R}} A + e^{\frac{i x^+}{R}} B + C + e^{-\frac{i x^+}{R}} D } \sum_{k\in\cal{S}^{(5d)}} \psi_k e^{-\frac{i k x^+}{R}}
\end{equation}
where $A,B,C,D$ are $x^+$-independent matrices whose form will not be important. It immediately follows from this that the set of allowed modes is just $\cal{S}^{(5d)} = \bb{Z}$.

\subsubsection{The Action for 3d}

We can do the exact same analysis for $D=4$. However, we now have
\begin{equation} \label{eq: omega gamma 4d}
    \Omega_{ij} \gamma_{ij} = \frac{2i}{R} \gamma_{01} \gamma_* \ ,
\end{equation}
so the action of the matrix on the field $\chi$ is
\begin{equation}
    \Omega_{ij} \gamma_{ij} \chi = \pm \frac{2i}{R} \chi \ ,
\end{equation}
depending on the chirality of our fields. Unlike in the 5d case no simplification occurs by choosing a particular sign; for definiteness we will take our fields to have positive chirality from here onwards. The effect of this is to shift $k$ to $k - \inv{4}$ in the action. For this reason it will be useful to redefine our field by an $x^+$-dependent phase, where we take
\begin{equation} \label{eq: spinor transformation 4d mod}
    \hat{\psi}(\hat{x}) = \Delta^{- \frac{3}{2}} S_+^{-1}[\Lambda] e^{- \frac{i x^+}{4R}} \psi(x)
\end{equation}
in place of \eqref{eq: spinor transformation} to simplify the action. With this, the action of the 3d theory is
\begin{equation} \label{eq: 4d action}
    S^{(3d)} = 2\pi i R \sum_{k\in\cal{S}^{(3d)}} \int d^3x \Bigg[ \sqrt{2} \lambda_k^{\dag} \partial_- \lambda_k + \chi_k^{\dag} \gamma_{0i} D_i \lambda_k + \lambda_k^{\dag} \gamma_{0i} D_i \chi_k - \frac{\sqrt{2} i k}{R}\chi_k^{\dag} \chi_k \Bigg] \ .
\end{equation}
Let us determine the set $\cal{S}^{(3d)}$. Examining the consistency condition \eqref{eq: consistency} using the explicit form \eqref{eq: S for 4d} for the spinor representation and the modified transformation \eqref{eq: spinor transformation 4d mod} leads to
\begin{equation}
    \sum_{l\in\bb{Z}} \psi_{l}' e^{- \frac{i l x^+}{R}} = \brac{ e^{\frac{i x^+}{R}} A + B + e^{-\frac{i x^+}{R}} C } \sum_{k\in\cal{S}^{(3d)}} \psi_k e^{-\frac{i k x^+}{R}}
\end{equation}
in terms of $x^+$-independent matrices $A,B,C$, so as for the 5d theory we have $\cal{S}^{(3d)} = \bb{Z}$.

\subsection{A Simplification for Non-Zero \texorpdfstring{$k$}{k}}

The action of the level-$k$ mode takes the form
\begin{equation} \label{eq: fermion level k action}
    S_k = 2\pi i R \int d^{2n-1}x \Bigg[ \sqrt{2} \lambda_k^{\dag} \partial_- \lambda_k + \chi_k^{\dag} \gamma_{0i} D_i \lambda_k + \lambda_k^{\dag} \gamma_{0i} D_i \chi_k - \frac{\sqrt{2}ik}{R} \chi_k^{\dag} \chi_k
    \Bigg]
\end{equation}
in both 3d and 5d. The field $\chi_k$ plays a remarkably different role depending on whether $k=0$ or $k\neq0$. In the former case, the final term in the action vanishes. The field $\chi_0$ is then a spinorial Lagrange multiplier imposing the constraint
\begin{equation}
    \gamma_{0i} D_i \lambda_k = 0 \ .
\end{equation}
However, when $k\neq0$ $\chi_k$ has the algebraic equation of motion
\begin{equation} \label{eq: chi eom non-zero k}
    \chi_k = - \frac{iR}{\sqrt{2}k} \gamma_{0i} D_i \lambda_k \ .
\end{equation}
We can substitute this back into the action to obtain
\begin{equation} \label{eq: lambda scalar-like action}
    S_k = \frac{\sqrt{2} \pi R^2}{k} \int d^{2n-1} x \Bigg[ \frac{2i}{R} \lambda_k^{\dag} \brac{k - \Gamma} \partial_- \lambda_k + \lambda_k^{\dag} D^2 \lambda_k
    \Bigg] \ ,
\end{equation}
where we've defined the matrix
\begin{equation}
    \Gamma = \frac{i R \Omega_{ij} \gamma_{ij}}{4} \ .
\end{equation}
As mentioned above, the properties of this matrix are dimension dependent. Using \eqref{eq: omega gamma 4d} we see that in 3d we have
\begin{equation}
    \Gamma \lambda_k = \inv{2} \lambda_k \ ,
\end{equation}
so the action is just
\begin{equation}
    S^{(3d)}_k = \frac{\sqrt{2} \pi R^2}{k} \int d^3 x \Bigg[ \frac{2i}{R} \brac{k - \inv{2}} \lambda_k^{\dag}  \partial_- \lambda_k + \lambda_k^{\dag} D^2 \lambda_k
    \Bigg] \ .
\end{equation}
In 5d it has the nice property that
\begin{equation}
    P_- \Gamma^2 = P_- \ ,
\end{equation}
so we can define the pseudo-projector
\begin{equation}
    P_{\pm}^{\Gamma} = \inv{2} \brac{ \bbm{1} \pm \Gamma }
\end{equation}
that behaves as a projector when acting on the subspace defined by projecting with $P_-$. Defining the fields
\begin{equation}
    \lambda_k^{\pm} = P^{\Gamma}_{\pm} \lambda_k \ ,
\end{equation}
the action decomposes as
\begin{equation}
    S^{(5d)}_k = \frac{\sqrt{2} \pi R^2}{k} \int d^5 x \Bigg[ \frac{2i(k-1)}{R} \lambda_k^{+\dag} \partial_- \lambda^+_k + \frac{2i(k+1)}{R} \lambda_k^{-\dag} \partial_- \lambda^-_k + \lambda_k^{+\dag} D^2 \lambda^+_k + \lambda_k^{- \dag} D^2 \lambda^-_k
    \Bigg] \ .
\end{equation}
By counting the degrees of freedom in our fields we observe that both $\lambda_k$ in the 4d case and $\lambda_k^{\pm}$ in the 6d case are complex Grassmann numbers. It is then no surprise that their actions are almost identical to the level-$k$ action for the $SU(1,n)$-invariant free complex scalar field
\begin{equation} \label{eq: phi action}
    S^{(\phi)}_k = 2\pi R \int d^{2n-1}x \Bigg[ \frac{2i k}{R} \Bar{\phi}_k \partial_- \phi_k + \Bar{\phi}_k D^2 \phi_k \Bigg] 
\end{equation}
obtained in \cite{Lambert:2021nol}. The novel part of this is the shift in $k$; we will see later when we examine the Ward identities satisfied by the two-point functions that this is due to the non-trivial transformations of the fields under spatial rotations. 

We can use this form of the action to predict the form of the $\lambda_k\lambda_k^{\dag}$ two-point function without doing any detailed computations. Defining the $(2n-1)$-dimensional scalar two-point function
\begin{align}
    G_k^{((2n-1)d)} \equiv \vac{\phi_k(x_1^-,x_1) \Bar{\phi}_k(x_2^-,x_2)}
\end{align}
for the action \eqref{eq: phi action}, matching the normalisation of the fields in the above actions suggests that the fermion two-point functions should be related to the scalar two-point function by
\begin{equation}
    \vac{\lambda_k(x_1^-,x_1) \lambda_k^{\dag}(x_2^-,x_2)} = \frac{\sqrt{2} k}{R} P_- G_{k-\inv{2}}^{(3d)}
\end{equation}
in 3d, and
\begin{equation}
    \vac{\lambda_k (x_1^-,x_1) \lambda_k^{\dag} (x_2^-,x_2) } = \frac{\sqrt{2} k}{R} P_- \brac{ P_+^{\Gamma} G_{k-1}^{(5d)} + P_-^{\Gamma} G_{k+1}^{(5d)}}
\end{equation}
in 5d. The projection operators have been added by hand to project onto the subspaces in which each field lives. We will show in the next section that this intuition is correct.

\section{SUSY and Correlation Functions} \label{sec: susy and corr}
\subsection{Supersymmetry in the Reduced Theory}
\label{sec: susy}

Since the fermionic theory we are considering is free, the quantities of interest are the two-point functions of the fields. There are many ways we could compute these, but it will prove convenient to fix these using the supersymmetry present in a free theory with a single complex scalar and fermion and the known two-point functions for an $SU(1,n)$ scalar field \cite{Lambert:2021nol}. 

We start with the action for a free theory in Minkowski spacetime consisting of complex scalar and Weyl spinor fields,
\begin{equation}
    S = - \int d^{2n} \hat{x} \brac{ \eta^{\mn} \hpart_{\mu} \Bar{\hat{\phi}} \hpart_{\nu} \hat{\phi} + \bar{\hat{\psi}} \gamma^{\mu}\hpart_{\mu} \psi} \ .
\end{equation}
This is invariant under the transformation
\begin{subequations}
\begin{gather}
    \delta \hat{\phi} = \Bar{\hat{\xi}} \hat{\psi} \ ,\\
    \delta \hat{\psi} = -\hpart_{\mu} \hat{\phi} \gamma^{\mu} \hat{\xi} \ ,
\end{gather}
\end{subequations}
where $\hat{\xi}$ is a Weyl spinor. We can then use the coordinate transformation \eqref{eq: coord transformation} and perform the reduction along $x^+$, leaving us with an infinite tower of scalar and spinor theories in $(2n-1)$-dimensions. Since we still have a full description of the system (albeit in terms of an infinite number of fields) after transforming and taking the null reduction, the supersymmetry of the $2n$-dimensional theory will be present in some form in the $(2n-1)$-dimensional theory. As the supercharge transforms as a $2n$-dimensional spinor, its components will generically develop dependence on $x^+$ and will therefore not commute with the generator of translations along the compactified null direction. This means that we expect some fraction of the supersymmetry transformations in the reduced theory will take us between levels of the Kaluza-Klein tower.

\subsubsection{3d Supersymmetry}
Let us first look at the 3d theory. Combining the scalar field action \eqref{eq: phi action} for $n=2$ and $k^{(\phi)}\in\bb{Z}+\inv{2}$ with the fermion action \eqref{eq: 4d action} gives
\begin{align}
    S = 2\pi R \sum_{k\in\bb{Z}} \int d^3x \Bigg[& \frac{2i}{R}\brac{k - \inv{2}} \bar{\phi}_{k-\inv{2}} \partial_- \phi_{k - \inv{2}} + \bar{\phi}_{k - \inv{2}} D^2 \phi_{k- \inv{2}} + i \sqrt{2} \lambda_k^{\dag} \partial_- \lambda_k \\ \nonumber
    &+ i \chi_k^{\dag} \gamma_{0i} D_i \lambda_k + i \lambda_k^{\dag} \gamma_{0i} D_i \chi_k + \frac{\sqrt{2} k}{R} \chi_k^{\dag} \chi_k
    \Bigg] \ .
\end{align}
We introduce a 4d Weyl spinor parameter $\epsilon$, from which we can define the $\gamma_{01}$ chiral components
\begin{subequations}
\begin{align}
    \xi &= P_{+} \epsilon \ , \\
    \eta &= P_- \epsilon \ .
\end{align}
\end{subequations}
A short computation shows that the transformations
\begin{subequations}
\begin{align}
    \delta \phi_{k-\inv{2}} &= \eta^{\dag} \lambda_k \ , \\ 
    \delta \lambda_k &= - \frac{\sqrt{2} k}{R} \phi_{k- \inv{2}} \eta \ , \\
    \delta \chi_k &= i \gamma_{0i} D_i \phi_{k- \inv{2}} \eta \ ,
\end{align}
\end{subequations}
and
\begin{subequations}
\begin{align}
    \delta \phi_{k+ \inv{2}} &= \xi^{\dag} \chi_k \ , \\
    \delta \lambda_k &= i \gamma_{0i} D_i \phi_{k+\inv{2}} \xi \ , \\
    \delta \chi_k &= -i \sqrt{2} \partial_- \phi_{k+\inv{2}} \xi \ ,
\end{align}
\end{subequations}
both leave the action invariant. The commutators of these transformations on the scalar modes are
\begin{subequations}
\begin{align}
    [\delta_{\eta_1}, \delta_{\eta_2}] \phi_{k- \inv{2}} &= \frac{\sqrt{2} k}{R} \brac{ \eta_1^{\dag} \eta_2 - \eta_2^{\dag} \eta_1 } \phi_{k- \inv{2}} \ , \\
    [\delta_{\xi_1}, \delta_{\xi_2}] \phi_{k- \inv{2}} &= i \sqrt{2} \brac{ \xi_1^{\dag} \xi_2 - \xi_2^{\dag} \xi_1 } \partial_-\phi_{k- \inv{2}} \ , \\
    [\delta_{\eta}, \delta_{\xi}] \phi_{k- \inv{2}} &= i \xi^{\dag} \gamma_{0i} \eta D_i \phi_{k- \frac{3}{2}} - i \eta^{\dag} \gamma_{0i} \xi D_i \phi_{k+ \frac{1}{2}} \ .
\end{align}
\end{subequations}
We observe that we have closure of the algebra for the commutators between transformations of the same chirality, but that this breaks down for commutators between transformations of mixed chirality. It naively appears that these transformations are not symmetries of the theory. However, since we have an infinite number of fields we can repackage them into a sum of the form
\begin{equation}
    \phi = \sum_{k\in \bb{Z}} e^{- \frac{i k \alpha}{R}} \phi_{k-\inv{2}}
\end{equation}
in terms of an arbitrary parameter $\alpha$. We also rescale our spinor parameters to
\begin{subequations} \label{eq: spinor parameter redef}
\begin{align}
    \hat{\eta} &= e^{ \frac{i \alpha(1+c)}{2R}} \eta \ , \\ 
    \hat{\xi} &= e^{-\frac{i \alpha(1-c)}{2R}} \xi \ ,
\end{align}
\end{subequations}
where $c$ is a real constant. The action of the commutators on the field $\phi$ using the rescaled spinor parameters is
\begin{subequations}
\begin{align}
    [\delta_{\hat{\eta}_1}, \delta_{\hat{\eta}_2}] \phi &= i \sqrt{2} \brac{ \eta_1^{\dag}  \eta_2 - \eta_2^{\dag} \eta_1 } \partial_{\alpha} \phi \ , \\
    [\delta_{\hat{\xi}_1}, \delta_{\hat{\xi}_2}] \phi &= i \sqrt{2} \brac{ \xi_1^{\dag} \xi_2 - \xi_2^{\dag} \xi_1 } \partial_-\phi \ , \\
    [\delta_{\hat{\eta}}, \delta_{\hat{\xi}}] \phi &= i \brac{\xi^{\dag} \gamma_{0i} \eta - \eta^{\dag} \gamma_{0i} \xi} D_i \phi \ .
\end{align}
\end{subequations}
As expected, the algebra closes on the new field. We are able to identify $\alpha$ with the coordinate of the emergent dimension. Note that for this argument to work it is crucial that we retain the full Kaluza-Klein tower of fields.

\subsubsection{5d Supersymmetry}

We can repeat the analysis performed above for the 5d theory. In this case, the combined action is
\begin{align}
    S = 2\pi R \sum_{k\in\bb{Z}} \int d^3x \Bigg[& \frac{2i k}{R} \bar{\phi}_{k} \partial_- \phi_{k} + \bar{\phi}_{k} D^2 \phi_{k} + i \sqrt{2} \lambda_k^{\dag} \partial_- \lambda_k \\ \nonumber
    &+ i \chi_k^{\dag} \gamma_{0i} D_i \lambda_k + i \lambda_k^{\dag} \gamma_{0i} D_i \chi_k + \frac{\sqrt{2} k}{R} \chi_k^{\dag} \chi_k
    \Bigg] \ .
\end{align}
Our supersymmetry parameter $\epsilon$ is a 6d Weyl spinor, which we split into the components $(\xi,\eta)$ defined by
\begin{subequations}
\begin{gather}
    \xi = P_+ \epsilon \ , \\
    \eta = P_- \epsilon \ .
\end{gather}
\end{subequations}
It will be necessary to further split $\eta$ as
\begin{equation}
    \eta_{\pm} = P_{\pm}^{\Gamma} \eta \ .
\end{equation}
It can then be shown that the transformations
\begin{subequations}
\begin{align}
    \delta \phi_k &= \xi^{\dag} \chi_k \ , \\
    \delta \lambda_k &= i \gamma_{0i} D_i \phi_k \xi \ ,\\
    \delta \chi_k &= - i \sqrt{2} \partial_- \phi_k \xi \ ,
\end{align}
\end{subequations}
and
\begin{subequations}
\begin{align}
    \delta \phi_k &= \eta_{\pm}^{\dag} \lambda_{k\pm 1} \ , \\
    \delta \lambda_k &= - \frac{\sqrt{2} k}{R} \phi_{k\mp 1} \eta_{\pm} \ , \\
    \delta \chi_k &= i \gamma_{0i} D_i \phi_{k\mp 1} \eta_{\pm} \ ,
\end{align}
\end{subequations}
leave the action invariant. The commutators of these transformations on $\phi_k$ are
\begin{subequations}
\begin{align}
    [\delta_{\xi_1}, \delta_{\xi_2}] \phi_k &= i \sqrt{2} \brac{ \xi_1^{\dag} \xi_2 - \xi_2^{\dag} \xi_1 } \partial_- \phi_k \ , \\
    [\delta_{\eta_{1\pm}}, \delta_{\eta_{2\pm}}] \phi_k &= \frac{\sqrt{2}(k\pm 1)}{R} \brac{ \eta_{1\pm}^{\dag} \eta_{2\pm} - \eta_{2\pm}^{\dag} \eta_{1\pm}} \phi_k \ , \\
    [\delta_{\eta_{1\pm}}, \delta_{\eta_{2\mp}}] \phi_k &= 0 \ , \\
    [\delta_{\xi}, \delta_{\eta_{\pm}}] \phi_k &= i \brac{ \eta_{\pm}^{\dag} \gamma_{0i} \xi D_i \phi_{k\pm1} - \xi^{\dag} \gamma_{0i} \eta_{\pm} D_i \phi_{k\mp 1} } \ .
\end{align}
\end{subequations}
As in the 3d case, we do not have closure of the algebra between transformations of mixed $\gamma_{01}$ chirality. However, if we again work with the sum
\begin{equation}
    \phi = \sum_{k\in \bb{Z}} e^{- \frac{i k \alpha}{R}} \phi_k
\end{equation}
and the rescaled spinor variables\footnote{We are again free to redefine all spinor variables by an additional factor like in \eqref{eq: spinor parameter redef}, though we will choose not to do this.}
\begin{subequations}
\begin{align}
    \hat{\xi} &= \xi \ , \\
    \hat{\eta}_{\pm} &= e^{\pm \frac{i \alpha}{R}} \eta_{\pm} \ ,
\end{align}
\end{subequations}
the commutators become
\begin{subequations}
\begin{align}
    [\delta_{\hat{\xi}_1}, \delta_{\hat{\xi}_2}] \phi &=  \sqrt{2} i \brac{ \xi_1^{\dag} \xi_2 - \xi_2^{\dag} \xi_1 } \partial_- \phi \ , \\
    [\delta_{\hat{\eta}_{1\pm}}, \delta_{\hat{\eta}_{2\pm}}] \phi &= \sqrt{2} i \brac{ \eta_{1\pm}^{\dag} \eta_{2\pm} - \eta_{2\pm}^{\dag} \eta_{1\pm}} \brac{\partial_{\alpha} \mp \frac{i}{R}} \phi \ , \\
    [\delta_{\hat{\eta}_{1\pm}}, \delta_{\hat{\eta}_{2\mp}}] \phi &= 0 \ , \\
    [\delta_{\hat{\xi}}, \delta_{\hat{\eta}_{\pm}}] \phi &= i \brac{ \eta_{\pm}^{\dag} \gamma_{0i} \xi  - \xi^{\dag} \gamma_{0i} \eta_{\pm} } D_i \phi \ ,
\end{align}
\end{subequations}
and we recover closure of the algebra.

\subsection{Correlation Functions in the Reduced Theory}
\label{sec: corr}
We would like to find the two-point functions of the fermionic theory. Up to a factor, these are solutions to the classical equation of motion and are not hard to find. However, we need to fix the overall normalisation. The simplest way to do this is to exploit the fact that the theory we're interested in can be extended to the supersymmetric theory discussed above, so the normalisation of the fermion two-point functions can be fixed in terms of the known normalisation of scalar correlation functions found in \cite{Lambert:2021nol}. A derivation of this normalisation, with discussion of how it relates to the Green's function equations of the reduced theory, is given in appendix \ref{sec: scalar coefficients}.

\subsubsection{3d Correlation Functions}

We first discuss the correlation functions of the 3d theory. As shown in appendix \ref{sec: scalar coefficients 3d}, the non-zero correlation functions of the modes of a complex scalar field are
\begin{equation}
    \bra{\Omega} \phi_k(x^-,x) \bar{\phi}_k(0) \ket{\Omega} = - \frac{i}{8\pi^2 R} \inv{ \sqrt{z \bar{z}}} \brac{\frac{z}{\bar{z}}}^k
\end{equation}
for $k>0$, and
\begin{equation}
    \bra{\Omega} \bar{\phi}_k(x^-,x) \phi_k(0) \ket{\Omega} =  \frac{i}{8\pi^2 R} \inv{ \sqrt{z \bar{z}}} \brac{\frac{z}{\bar{z}}}^{-k}
\end{equation}
for $k<0$, where we have defined the complex variable $z$ as
\begin{equation} \label{eq: z def}
    z = x^- + \frac{i x^2}{4R} \ .
\end{equation}
As our theory is invariant under the action of the supercharges $Q_+$ and $Q_-$, the vacuum $\ket{\Omega}$ satisfies
\begin{equation}
    Q_+ \ket{\Omega} = Q_- \ket{\Omega} = Q_+^{\dag} \ket{\Omega} = Q_-^{\dag} \ket{\Omega} = 0 \ .
\end{equation}
This implies the correlation function identity
\begin{equation}
    \bra{\Omega} \{ Q_- , \phi_{k- \inv{2}}(x^-,x) \lambda_k^{\dag}(0) \} \ket{\Omega} = 0 \ ,
\end{equation}
which after expanding and simplifying gives us the fermion two-point function
\begin{align} \nonumber
    \bra{\Omega} \lambda_k(x^-,x) \lambda_k^{\dag}(0) \ket{\Omega} &= \frac{\sqrt{2} k}{R} P_- \bra{\Omega} \phi_{k-\inv{2}}(x^-,x) \bar{\phi}_{k-\inv{2}}(0) \ket{\Omega} \\ 
     &= - \frac{i k}{4\sqrt{2} \pi^2 R^2} \inv{\sqrt{z^2}} \brac{\frac{z}{\bar{z}}}^{k} P_-
\end{align}
for $k>0$ and zero for $k$ outside this range. Proceeding similarly for other combinations of supercharges acting on fields gives the non-vanishing two-point functions
\begin{subequations}
\begin{align}
    \bra{\Omega} \lambda_k(x^-,x) \chi_k^{\dag}(0) \ket{\Omega} &= -\frac{ik}{8\pi^2 R^2} \frac{\gamma_{0i} x^i}{z \sqrt{\bar{z}^2}} \brac{\frac{z}{\bar{z}}}^k  P_+  \ , \\
    \bra{\Omega} \chi_k(x^-,x) \lambda_k^{\dag}(0) \ket{\Omega} &= -\frac{ik}{8\pi^2 R^2} \frac{\gamma_{0i} x^i}{\sqrt{z^2} \bar{z}} \brac{\frac{z}{\bar{z}}}^k   P_- \ , \\
    \bra{\Omega} \chi_k(x^-,x) \chi_k^{\dag}(0) \ket{\Omega} &= \inv{4\sqrt{2} \pi^2 R} \inv{z \sqrt{\bar{z}^2}} \brac{\frac{z}{\bar{z}}}^k \brac{k - (k+1) \frac{z}{\bar{z}}} P_+ \ ,
\end{align}
\end{subequations}
for $k \geq 0 $, and
\begin{subequations}
\begin{align}
    \bra{\Omega} \lambda_k^{\dag}(x^-,x) \lambda_k(0) \ket{\Omega} &= -\frac{i k}{4\sqrt{2} \pi^2 R^2} \inv{\sqrt{\bar{z}^2}} \brac{\frac{z}{\bar{z}}}^{-k} P_- \ , \\
    \bra{\Omega} \lambda_k^{\dag}(x^-,x) \chi_k(0) \ket{\Omega} &= -\frac{ik}{8\pi^2 R^2} \frac{\gamma_{0i} x^i }{\sqrt{z^2} \bar{z}} \brac{\frac{z}{\bar{z}}}^{-k} P_- \ , \\
    \bra{\Omega} \chi_k^{\dag}(x^-,x) \lambda_k(0) \ket{\Omega} &= -\frac{ik}{8\pi^2 R^2} \frac{\gamma_{0i} x^i }{z \sqrt{\bar{z}^2}} \brac{\frac{z}{\bar{z}}}^{-k} P_+ \ , \\
    \bra{\Omega} \chi_k^{\dag}(x^-,x) \chi_k(0) \ket{\Omega} &= -\inv{4\sqrt{2} \pi^2 R} \inv{\sqrt{z^2} \bar{z}} \brac{\frac{z}{\bar{z}}}^{-k} \brac{k - (k+1) \frac{\bar{z}}{z}} P_+ \ ,
\end{align}
\end{subequations}
for $k\leq -1$. We note that the correlation functions should always read in matrix form, i.e.
\begin{equation}
    \bra{\Omega} (\lambda_k^{\dag})^{\beta} \brac{\chi_k}_{\alpha} \ket{\Omega} \equiv \tensor{\brac{ \bra{\Omega} \lambda_k^{\dag} \chi_k \ket{\Omega}}}{_{\alpha}^{\beta}} \ .
\end{equation}

As we have found the two-point functions of the modes, we can now resum them to obtain the two-point function of the original theory. We will be considering time-ordered two-point functions, where the time-ordering is taken with respect to the coordinate $x^-$. The two-point function of the resummed field $\lambda$ is given by
\begin{align} \nonumber
    \bra{\Omega} T\{ \lambda(x^+,x^-,x) \lambda^{\dag}(0) \} \ket{\Omega} &= \sum_{k\in \bb{Z}} e^{- \frac{i k x^+}{R}} \bra{\Omega} T\{ \lambda_k(x^-,x) \lambda_k^{\dag}(0) \} \ket{\Omega} \\  \nonumber
    &= \sum_{k=1}^{\infty} \bigg( e^{- \frac{ik x^+}{R}} \Theta(x^-) \bra{\Omega} \lambda_k(x^-,x) \lambda_k^{\dag}(0) \ket{\Omega} \\ 
    & \hspace{1.5cm} - e^{ \frac{ik x^+}{R}} \Theta(-x^-) \bra{\Omega}  \lambda_{-k}^{\dag}(-x^-,-x) \lambda_{-k}(0) \ket{\Omega} \bigg) \ .
\end{align}
We have used translational invariance of the vacuum to move all coordinate dependence to the first field. Substituting in the mode correlation functions and taking $\sqrt{z^2} = z$ gives
\begin{equation} \hspace{-1.0cm}
    \bra{\Omega} T\{ \lambda(x^+,x^-,x) \lambda^{\dag}(0) \} \ket{\Omega} = - \frac{i P_-}{4\sqrt{2} \pi^2 R^2} \inv{z} \sum_{k=1}^{\infty} \brac{ \Theta(x^-) k \brac{\frac{z e^{- \frac{i x^+}{R}}}{\bar{z}}}^k - \Theta(-x^-) k \brac{\frac{z e^{- \frac{i x^+}{R}}}{\bar{z}}}^{-k} } \ .
\end{equation}
However, these sums do not converge: we need to specify an $i\epsilon$ prescription. As $x^-$ has a definite sign in each term, we can take
\begin{equation} \label{eq: i epsilon}
    x^+ \to x^+ - i x^- \epsilon \equiv x^+_{\epsilon}
\end{equation}
for some infinitesimal $\epsilon>0$. We will explain the relation of this to the usual $i\epsilon$ prescription below. The sums are now convergent and (suppressing the $\epsilon$ terms for now) we find the result
\begin{equation}
    \bra{\Omega} T\{ \lambda(x^+,x^-,x) \lambda^{\dag}(0) \} \ket{\Omega} = - \frac{i P_-}{4\sqrt{2} \pi^2 R^2} \frac{\bar{z}}{\brac{ \bar{z} e^{\frac{i x^+}{2R}} - z e^{-\frac{i x^+}{2R}} }^2} \ .
\end{equation}
The same manipulations for our other two-point functions lead to the resummed functions
\begin{subequations}
\begin{align}
    \bra{\Omega} T\{ \lambda(x^+,x^-,x) \chi^{\dag}(0) \} \ket{\Omega} &= -\frac{i P_-}{8\pi^2 R^2} \frac{\gamma_{0i} x^i}{\brac{ \bar{z} e^{\frac{i x^+}{2R}} - z e^{-\frac{i x^+}{2R}} }^2} \ , \\
    \bra{\Omega} T\{ \chi(x^+,x^-,x) \lambda^{\dag}(0) \} \ket{\Omega} &= -\frac{i P_+}{8\pi^2 R^2} \frac{\gamma_{0i} x^i}{\brac{ \bar{z} e^{\frac{i x^+}{2R}} - z e^{-\frac{i x^+}{2R}} }^2} \ , \\
    \bra{\Omega} T\{ \chi(x^+,x^-,x) \chi^{\dag}(0) \} \ket{\Omega} &= -\frac{i P_+}{2 \sqrt{2}\pi^2 R} \frac{e^{\frac{ix^+}{2R}} \sin \brac{\frac{x^+}{2R}}}{\brac{ \bar{z} e^{\frac{i x^+}{2R}} - z e^{-\frac{i x^+}{2R}} }^2} \ .
\end{align}
\end{subequations}
This is equivalent to the correlation function
\begin{equation} \label{eq: untransformed correlation function} \hspace{-1.5cm} 
    \bra{\Omega} T\{ \psi(x^+,x^-,x) \psi^{\dag}(0) \} \ket{\Omega} = - \frac{i}{8\pi^2 R^2} \inv{\brac{ \bar{z} e^{\frac{i x^+}{2R}} - z e^{-\frac{i x^+}{2R}} }^2} \brac{ 2\sqrt{2} R e^{\frac{ix^+}{2R}} \sin \brac{\frac{x^+}{2R}} P_+ + \sqrt{2} \bar{z} P_- + \gamma_{0i} x^i }
\end{equation}
for the full spinor field $\psi$. Recalling the transformation\footnote{Note that we should replace $x^+$ with $x_{\epsilon}^+$ in the transformation.} \eqref{eq: spinor transformation 4d mod}, the two-point function of the original Minkowski theory can be found from
\begin{align}
    \bra{\Omega} T\{ \hat{\psi}(\hat{x}^+,\hat{x}^-,\hat{x}) \hat{\psi}^{\dag}(0) \} \ket{\Omega} = e^{- \frac{i x^+}{4R}} \Delta^{-\frac{3}{2}}(x) S_+^{-1}[\Lambda_x] \bra{\Omega} T\{ \psi(x^+,x^-,x) \psi^{\dag}(0) \} \ket{\Omega} \ ,
\end{align}
which after evaluating and using
\begin{equation}
    \frac{\cos\brac{\frac{x^+}{2R}}}{z e^{- \frac{i x^+}{2R}} - \bar{z} e^{ \frac{i x^+}{2R}}} = \frac{- 2i R}{- 2 \hat{x}^+ \hat{x}^- + \hat{x}^i \hat{x}^i}
\end{equation}
gives
\begin{align} \label{eq: 4d normalised minkowski 2 point}\nonumber
    \bra{\Omega} T\{ \hat{\psi}(\hat{x}^+,\hat{x}^-,\hat{x}) \hat{\psi}^{\dag}(0) \} \ket{\Omega} &= - \frac{i \cos^2\brac{\frac{x^+}{2R}}}{8\pi^2 R^2} \frac{\sqrt{2} \hat{x}^+ P_+ + \sqrt{2} \hat{x}^- P_- + \gamma_{0i} \hat{x}^i}{\brac{ \bar{z} e^{\frac{i x^+}{2R}} - z e^{-\frac{i x^+}{2R}} }^2} \\
    &= -\frac{i}{2\pi^2} \frac{\gamma_{\mu} \gamma_0 \hat{x}^{\mu} }{\brac{-2 \hat{x}^+ \hat{x}^- + \hat{x}^i \hat{x}^i}^2} \ ,
\end{align}
or equivalently
\begin{equation}
    \bra{\Omega} T\{ \hat{\psi}(\hat{x}^+,\hat{x}^-,\hat{x}) \hat{\bar{\psi}}(0) \} \ket{\Omega} = \frac{\gamma_{\mu} \hat{x}^{\mu} }{2\pi^2 \brac{-2 \hat{x}^+ \hat{x}^- + \hat{x}^i \hat{x}^i}^2} \ .
\end{equation}
We recognise this as the two-point correlation for a free fermion in 4d Minkowski spacetime when written in lightcone coordinates, with the canonically normalised action
\begin{equation}
    S = -\int d^4\hat{x} \, \hat{\bar{\psi}} \gamma^{\mu} \partial_{\mu} \hat{\psi} \ .
\end{equation}
We should justify the strange coordinate-dependent $i\epsilon$ prescription used to ensure convergence of the sum over modes. We recover $\epsilon$ in our final correlation function by taking $\hat{x}^{\mu}$ to $\hat{x}^{\mu}_{\epsilon}$, where $\hat{x}^{\mu}_{\epsilon}$ is the coordinate transformation \eqref{eq: coord transformation} after replacing $x^+$ with $x^+_{\epsilon}$. If we examine the denominator of the correlation function as $R\to \infty$, we see that
\begin{align} \nonumber
    \lim_{R\to\infty}\brac{\inv{(-2 \hat{x}^+_{\epsilon} \hat{x}^-_{\epsilon} + \hat{x}^i_{\epsilon} \hat{x}^i_{\epsilon})^2} } & = \inv{\brac{ - 2 \brac{x^+ - i x^- \epsilon} x^- + x^i x^i}^2} \\
    &= \inv{ \brac{ - 2 x^+ x^- + x^i x^i + i \epsilon' }^2 } \ ,
\end{align}
where we've defined
\begin{equation}
    \epsilon' = 2 (x^-)^2 \epsilon \ .
\end{equation}
We recognise this as the standard $i\epsilon$ prescription for a time-ordered two-point function, justifying our procedure\footnote{Note that we should really redefine our final expression in terms of a constant infinitesimal $\epsilon'$ so that the prescription is also valid when $x^-=0$.}. We can also consider the resummed correlation function \eqref{eq: untransformed correlation function} in the transformed theory. Replacing $x^+$ by $x^+_{\epsilon}$ and expanding the denominator to first order in $\epsilon$ gives
\begin{equation}
    \vac{T\{\psi(x^+,x^-,x) \psi^{\dag}(0)\}} =  \frac{i \brac{2\sqrt{2} R e^{ \frac{i x^+}{2R}} \sin\brac{\frac{x^+}{2R}} P_+ + \sqrt{2} \Bar{z} P_- + \gamma_{0i} x^i }}{ 2\pi^2 \brac{x^2 \cos\brac{\frac{x^+}{2R}} - 4R x^- \sin\brac{\frac{x^+}{2R}} + i \epsilon'}^2} \ ,
\end{equation}
so we see that this is the natural $i\epsilon$ prescription in the theory with action \eqref{eq: action 1}. Further motivation can be found from the analogous computation in 2d, which is performed in section \ref{sec: 2d ferm reduction}, where the same prescription is found by ensuring integrals over momenta are well-defined.

\subsubsection{5d Correlation Functions}
The same procedure can be used to find the correlation functions of the 5d theory. As shown in appendix \ref{sec: scalar coefficients 5d}, the non-zero scalar two-point functions are
\begin{equation}
    \bra{\Omega} \phi_k(x^-,x) \bar{\phi}_k(0) \ket{\Omega} = - \frac{k}{16\pi^3 R^2} \inv{z \bar{z}} \brac{\frac{z}{\bar{z}}}^k
\end{equation}
for $k>0$, and
\begin{equation}
    \bra{\Omega} \bar{\phi}_k(x^-,x) \phi_k(0) \ket{\Omega} = \frac{k}{16\pi^3 R^2} \inv{z \bar{z}} \brac{\frac{z}{\bar{z}}}^{-k}
\end{equation}
for $k<0$. The only difference between this and the 3d case is that, in 5d, the action of $Q_-$ decomposes into the two subspaces defined by $\Gamma$, with the value of $k$ shifting in opposite directions depending on the subspace. This requires that the correlation function identities involving $Q_-$ contain two scalar fields in order to not project onto either subspace. For example, to get the $\lambda_k \lambda_k^{\dag}$ two-point function we must consider the identity
\begin{equation}
    \bra{\Omega} \{ Q_-, (\phi_{k+1} + \phi_{k-1}) \lambda_k^{\dag} \} \ket{\Omega} = 0 \ ;
\end{equation}
from this we find
\begin{equation} \hspace{-1.0cm}
    \bra{\Omega} \lambda_k(x^-,x) \lambda_k^{\dag}(0) \ket{\Omega} = \frac{\sqrt{2}k}{R} P_- \Big( P_+^{\Gamma} \bra{\Omega} \phi_{k-1}(x^-,x) \bar{\phi}_{k-1}(0) \ket{\Omega} + P_-^{\Gamma} \bra{\Omega} \phi_{k+1}(x^-,x) \bar{\phi}_{k+1}(0) \ket{\Omega} \Big) \ ,
\end{equation}
which after substituting in the appropriate scalar two-point functions is
\begin{equation}
    \bra{\Omega} \lambda_k(x^-,x) \lambda_k^{\dag} (0) \ket{\Omega} = - \frac{k}{8\sqrt{2} \pi^3 R^3} P_- \brac{\frac{z}{\bar{z}}}^k \brac{ \frac{(k+1) P_-^{\Gamma}}{\bar{z}^2} + \frac{(k-1) P_+^{\Gamma}}{z^2} }
\end{equation}
for $k>0$ and zero otherwise.

Proceeding similarly, the other non-vanishing two-point functions of the modes are
\begin{subequations}
\begin{align} \nonumber
    \vac{\lambda_k(x^-,x) \chi_k^{\dag}(0)} &= \frac{ik}{32\pi^3 R^3} \gamma_{0i} P_+ \brac{\inv{z \bar{z}}}^2 \brac{\frac{z}{\bar{z}}}^k \Bigg[ ix^i \Big( z(1+k) - \bar{z} (1-k) \Big) \\
    & \hspace{3.9cm} + R \Omega_{ij} x^j \Big( z(1+k) + \bar{z} (1-k) \Big) \Bigg] \ ,
    \\ \nonumber
    \vac{\chi_k(x^-,x) \lambda_k^{\dag}(0)} &= \frac{ik}{32\pi^3 R^3} \gamma_{0i} P_- \brac{\inv{z \bar{z}}}^2 \brac{\frac{z}{\bar{z}}}^k \Bigg[ ix^i \Big( z(1+k) - \bar{z} (1-k) \Big) \\
    & \hspace{3.9cm} - R \Omega_{ij} x^j \Big( z(1+k) + \bar{z} (1-k) \Big) \Bigg] \ , \\
    \vac{\chi_k(x^-,x) \chi_k^{\dag}(0)} &= \frac{i k}{8\sqrt{2} \pi^3 R^2 } P_+ \brac{\inv{z \bar{z}}}^2 \brac{ \frac{z}{\bar{z}}}^k \Big( (1+k) z + (1-k) \bar{z} \Big) \ ,
\end{align}
\end{subequations}
for $k>0$, and
\begin{subequations}
\begin{align} 
    \bra{\Omega} \lambda_k^{\dag}(x^-,x) \lambda_k(0) \ket{\Omega} &= - \frac{k}{8\sqrt{2} \pi^3 R^3} P_- \brac{\frac{z}{\bar{z}}}^{-k} \brac{ \frac{(k+1) P_-^{\Gamma}}{z^2} + \frac{(k-1) P_+^{\Gamma}}{\bar{z}^2}} \ , \\ \nonumber
    \vac{ \chi_k^{\dag}(x^-,x) \lambda_k(0)} &= \frac{ik}{32\pi^3 R^3} \gamma_{0i} P_+ \brac{\inv{z \bar{z}}}^2 \brac{\frac{z}{\bar{z}}}^{-k} \Bigg[ ix^i \Big( \bar{z}(1+k) - z (1-k) \Big) \\
    & \hspace{3.9cm} + R \Omega_{ij} x^j \Big( \bar{z}(1+k) + z (1-k) \Big) \Bigg] \ , \\ \nonumber
    \vac{ \lambda_k^{\dag}(x^-,x) \chi_k(0)} &= \frac{ik}{32\pi^3 R^3} \gamma_{0i} P_- \brac{\inv{z \bar{z}}}^2 \brac{\frac{z}{\bar{z}}}^{-k} \Bigg[ ix^i \Big( \bar{z}(1+k) - z (1-k) \Big) \\
    & \hspace{3.9cm} - R \Omega_{ij} x^j \Big( \bar{z}(1+k) + z (1-k) \Big) \Bigg] \ , \\
    \vac{\chi_k^{\dag}(x^-,x) \chi_k(0)} &= - \frac{ik}{8\sqrt{2}\pi^3 R^2} P_+ \brac{\inv{z \bar{z}}}^2 \brac{\frac{z}{\bar{z}}}^{-k} \Big( (1+k) \bar{z} + (1-k) z \Big) \ ,
\end{align}
\end{subequations}
for $k<0$.

We can resum the mode two-point functions using the same $i\epsilon$ prescription outlined above to obtain the two-point functions of the 6d fields. A short computation gives
\begin{subequations}
\begin{gather}
    \bra{\Omega} T\{ \lambda(x^+,x^-,x) \lambda^{\dag}(0)\} \ket{\Omega} = \frac{P_-}{4\sqrt{2} \pi^3 R^3} \frac{\cos\brac{\frac{x^+}{2R}} \brac{x^- - \frac{i x^2}{4R} \Gamma} - \sin\brac{\frac{x^+}{2R}} \brac{ix^- \Gamma + \frac{x^2}{4R}}}{\brac{z e^{- \frac{i x^+}{2R}} - \bar{z} e^{\frac{i x^+}{2R}}}^3} \ , \\
    \vac{T\{ \lambda(x^+,x^-,x) \chi^{\dag}(0) \} } = \frac{\gamma_{0i} P_+}{8\pi^3 R^3} \frac{x^i \cos\brac{\frac{x^+}{2R}} + R \Omega_{ij} x^j \sin\brac{\frac{x^+}{2R}}}{\brac{z e^{- \frac{i x^+}{2R}} - \bar{z} e^{\frac{i x^+}{2R}}}^3} \ , \\
    \vac{T\{ \chi(x^+,x^-,x) \lambda^{\dag}(0) \} } = \frac{\gamma_{0i} P_-}{8\pi^3 R^3} \frac{x^i \cos\brac{\frac{x^+}{2R}} - R \Omega_{ij} x^j \sin\brac{\frac{x^+}{2R}}}{\brac{z e^{- \frac{i x^+}{2R}} - \bar{z} e^{\frac{i x^+}{2R}}}^3} \ , \\ 
    \vac{T\{ \chi(x^+,x^-,x) \chi^{\dag}(0) \} } = \frac{P_+}{2\sqrt{2} \pi^3 R^2} \frac{\sin \brac{\frac{x^+}{2R}} }{{\brac{z e^{- \frac{i x^+}{2R}} - \bar{z} e^{\frac{i x^+}{2R}}}^3}} \ .
\end{gather}
\end{subequations}
We can combine these into the natural 6d field $\psi$, yielding the two-point function
\begin{align} \hspace{-1.1cm} \nonumber
    \vac{T\{\psi(x^+,x^-,x) \psi^{\dag}(0)\}} = \inv{8\pi^3 R^3} &\inv{\brac{z e^{- \frac{i x^+}{2R}} - \bar{z} e^{\frac{i x^+}{2R}}}^3} \Bigg[ \cos\brac{\frac{x^+}{2R}} \bigg( \sqrt{2} P_- \brac{x^- - \frac{i x^2}{4R} \Gamma} + \gamma_{0i} x^i\bigg) \\
    &+ \sin \brac{\frac{x^+}{2R}} \bigg( 2\sqrt{2} R P_+ - \sqrt{2} P_- \brac{i x^- \Gamma + \frac{x^2}{4R}} - R \Omega_{ij} \gamma_{1i} x^j \bigg) \Bigg] \ .
\end{align}
Using the transformation \eqref{eq: spinor transformation} we find that the original Minkowski field $\hat{\psi}$ has the two-point function
\begin{align} \nonumber
    \vac{T\{\hat{\psi}(\hat{x}^+,\hat{x}^-,\hat{x}) \hat{\psi}^{\dag}(0)\}} &= \Delta^{-\frac{5}{2}}(x) S^{-1}[\Lambda_x] \vac{T\{\psi(x^+,x^-,x) \psi^{\dag}(0)\}} \\ \nonumber
    &= \inv{8\pi^3 R^3} \frac{\cos^3\brac{\frac{x^+}{2R}}}{\brac{z e^{- \frac{i x^+}{2R}} - \bar{z} e^{\frac{i x^+}{2R}}}^3} \Bigg[ 2\sqrt{2} R \tan\brac{\frac{x^+}{2R}} P_+ \\
    & + \sqrt{2} \brac{x^- + \frac{x^2}{4R} \tan\brac{\frac{x^+}{2R}} } P_- + \gamma_{0i} \brac{x^i - \tan\brac{\frac{x^+}{2R}} R\Omega_{ij} x^j} \Bigg]
\end{align}
which after tidying up gives
\begin{equation} \label{eq: 6d normalised minkowski 2 point}
    \vac{T\{\hat{\psi}(\hat{x}^+,\hat{x}^-,\hat{x}) \hat{\psi}^{\dag}(0)\}} = -\frac{i}{\pi^3} \frac{\gamma_{\mu} \gamma_{0} \hat{x}^{\mu}}{\brac{- 2 \hat{x}^+ \hat{x}^- + \hat{x}^i \hat{x}^i}^3} \ ,
\end{equation}
or equivalently
\begin{equation}
    \vac{T\{\hat{\psi}(\hat{x}^+,\hat{x}^-,\hat{x}) \bar{\hat{\psi}}(0)\}} = \frac{1}{\pi^3} \frac{\gamma_{\mu} \hat{x}^{\mu}}{\brac{- 2 \hat{x}^+ \hat{x}^- + \hat{x}^i \hat{x}^i}^3} \ .
\end{equation}
As in the 4d case, this is the usual two-point function of a Minkowski spinor field with canonically normalised action
\begin{equation}
    S = - \int d^6\hat{x} \, \bar{\hat{\psi}} \gamma^{\mu} \partial_{\mu} \hat{\psi} \ .
\end{equation}

\section{Limits of \texorpdfstring{$SU(1,n)$}{SU(1,n)} Fermions} \label{sec: limits}
\subsection{The DLCQ Limit of \texorpdfstring{$SU(1,n)$}{SU(1,n)} Fermions}
\label{sec: DLCQ}

One of the novel properties of the $SU(1,n)$-invariant theories we have analysed is that they contain the DLCQ of the original Minkowski theory as a limiting case. We can find this using the procedure described in \cite{Lambert:2020zdc}, which we briefly review. Recall that we fixed the moding of theory by asking that the 'physical' fields\footnote{I.e. those obtained from a diffeomorphism of the original Minkowski fields.} admitted a Fourier series expansion on the interval $x^+\in[-\pi R, \pi R]$. If we choose some $N\in\bb{N}$ we can instead ask that the physical fields are periodic with period $2\pi R_+$, where we define $R_+$ as
\begin{equation} \label{eq: R+ def}
    R_+ = \frac{R}{N} \ .
\end{equation}
In the fermionic case of interest the moding of both the physical and Weyl-transformed fields matched, so the transformed fields will also be periodic under a $2\pi R_+$ shift in $x^+$. The set of allowed modes for our fields is restricted to $\cal{S} = N\bb{Z}$; in other words, we can express each mode number $k$ as
\begin{equation} \label{eq: k+ def}
    k = k_+ N
\end{equation}
for some $k_+\in\bb{Z}$. This means the Kaluza-Klein tower takes the form
\begin{equation}
    \psi(x^+,x^-,x) = \sum_{k_+\in\bb{Z}} e^{- \frac{i k_+}{R_+}} \psi_{k_+}(x^-,x) \ .
\end{equation}
As taking the limit $N\to\infty$ with $R_+$ fixed sends $R\to\infty$ the coordinate transformation \eqref{eq: coord transformation} is trivial and the $\Omega$-deformation of the spacetime metric disappears. This leaves us with a field theory on flat Minkowski spacetime where our fields are periodic along a finite null interval: we recognise this as the DLCQ of the theory.

There are two ways to find the two-point functions of the DLCQ theory. We can either start with the theory in the DLCQ limit and find its correlation functions, or we can directly take the limit of the $SU(1,n)$ fermionic two-point functions. Both methods will produce the same results\footnote{For a more detailed discussion of this point for a scalar field theory see appendix \ref{sec: appendix DLCQ}.}, so for convenience we will work with the latter approach. Let us consider the action \eqref{eq: fermion level k action} for a single level of the reduced theory. Making the replacements \eqref{eq: R+ def} and \eqref{eq: k+ def}, we see that we need to rescale our fields to
\begin{subequations}
\begin{align}
    \lambda_k(x^-,x) &= \inv{\sqrt{N}} \lambda_{k_+}(x^-,x) \ , \\
    \chi_k(x^-,x) &= \inv{\sqrt{N}} \chi_{k_+}(x^-,x) \ ,
\end{align}
\end{subequations}
for the $N\to\infty$ limit to be well-defined. With this rescaling the action becomes
\begin{equation}
    S_k \to 2\pi R \int d^{2n-1}x \brac{\sqrt{2} \lambda_k^{\dag} \partial_- \lambda_k + \chi^{\dag} \gamma_{0i} \partial_i \lambda_k + \lambda_k^{\dag} \gamma_{0i} \partial_i \chi_k - \frac{\sqrt{2} i k}{R} \chi_k^{\dag} \chi_k} 
\end{equation}
as we take $N\to\infty$, where we have dropped all pluses for convenience. We note in passing that this is invariant under the Schr\"odinger group.

Applying this limit to the 3d and 5d two-point functions gives
\begin{subequations}
\begin{align}
    \vac{T \{ \lambda_k(x^-,x) \lambda_k^{\dag}(0)\}} &= \frac{\sqrt{2} \, \cal{T}^{(n)}_k  P_-}{4\pi R} \ , \\
    \vac{T \{ \lambda_k(x^-,x) \chi_k^{\dag}(0) \} } &= \frac{\cal{T}^{(n)}_k \gamma_{0i} x^i P_+}{4\pi R\, x^-} \ , \\
    \vac{T \{ \chi_k(x^-,x) \lambda_k^{\dag}(0) \} } &= \frac{\cal{T}^{(n)}_k \gamma_{0i} x^i P_-}{4\pi R \, x^-} \ , \\
    \vac{T \{ \chi_k(x^-,x) \chi_k^{\dag}(0) \} } &= 
    \frac{\sqrt{2}P_+ \brac{\pi \cal{T}_k^{(n)} x^2 - (n-1) \cal{T}_k^{(n-1)} }}{8\pi^2 R \, (x^-)^2} \ ,
\end{align}
\end{subequations}
where we've defined
\begin{equation}
    \cal{T}_k^{(\alpha)} =  \begin{cases}
        \Theta(x^-) \brac{\frac{-i k}{2\pi R \, x^-}}^{\alpha-1} \exp\brac{\frac{i k x^2}{2R\, x^-}} & k>0 \\
        0 & k=0 \\
        -\Theta(-x^-) \brac{\frac{-i k}{2\pi R \, x^-}}^{\alpha-1} \exp\brac{\frac{i k x^2}{2R\, x^-}} & k < 0
    \end{cases} \ .
\end{equation}

We can resum these to obtain the two-point function of a free massless fermion on Minkowski spacetime with a periodic null direction. We have
\begin{align} \nonumber
    \sum_{k\in\bb{Z}} e^{-\frac{ik x^+}{R}} \cal{T}_k^{(\alpha)} &= \brac{\frac{-i}{2\pi R \, x^-}}^{\alpha-1} \sum_{k=1}^{\infty} \brac{ \Theta(x^-) k^{\alpha-1} q^k  + (-1)^{\alpha} \Theta(-x^-) k^{\alpha-1} q^{-k}  } \\
    &= \brac{\frac{-i}{2\pi R \, x^-}}^{\alpha-1} \Big(\Theta(x^-) \polylog{1-\alpha}{q} + (-1)^{\alpha} \Theta(-x^-) \polylog{1-\alpha}{q^{-1}} \Big)
\end{align}
in terms of the polylogarithm
\begin{equation} \label{eq: polylog def}
    \polylog{s}{z} = \sum_{k=1}^{\infty} \frac{z^k}{k^s} \ ,
\end{equation}
where we have defined
\begin{equation} \label{eq: q def}
    q = \exp\brac{ \frac{i}{2 R x^-} \brac{-2x^+ x^- + x^2}} \equiv \exp\brac{\frac{i x_{\mu} x^{\mu}}{2R \, x^-}}
\end{equation}
and used the usual $i\epsilon$ prescription \eqref{eq: i epsilon} to ensure convergence of the sum. When $\alpha\geq2$ we can use the identity
\begin{equation} \label{eq: polylog identity}
    \polylog{-s}{z} = (-1)^{s-1} \polylog{-s}{z^{-1}}
\end{equation}
valid for $s\in\bb{N}$ to simplify this to
\begin{equation}
    \sum_{k\in\bb{Z}} e^{-\frac{ik x^+}{R}} \cal{T}_k^{(\alpha)} = \brac{\frac{-i}{2\pi R \, x^-}}^{\alpha-1} \polylog{1-\alpha}{q} \ .
\end{equation}
We will also require the $\alpha=1$ case; we can then use the explicit form
\begin{equation}
    \polylog{0}{q} = \frac{q}{1-q}
\end{equation}
to obtain
\begin{equation}
    \sum_{k\in\bb{Z}} e^{- \frac{i kx^+}{R}} \cal{T}_k^{(1)} = \frac{i}{2} \brac{\cot\brac{\frac{x_{\mu} x^{\mu}}{4R \, x^-}} + i \big[\Theta(x^-) - \Theta(-x^-)\big] } \ .
\end{equation}
With these results in hand, we find the first three resummed two-point functions are
\begin{subequations}
\begin{align}
    \vac{T\{ \lambda(x^+, x^-,x) \lambda^{\dag}(0) \} } &= \frac{\sqrt{2} P_-}{4\pi R} \brac{\frac{-i}{2\pi R \, x^-}}^{n-1} \polylog{1-n}{q} \ , \\
    \vac{T \{ \lambda(x^+,x^-,x) \chi^{\dag}(0) \} } &= 
    \frac{\gamma_{0i} x^i P_+}{4\pi R\, x^-} \brac{\frac{-i}{2\pi R \, x^-}}^{n-1}  \polylog{1-n}{q} \ , \\
    \vac{T \{ \chi(x^+,x^-,x) \lambda^{\dag}(0) \} } &= 
    \frac{\gamma_{0i} x^i P_-}{4\pi R\, x^-} \brac{\frac{-i}{2\pi R \, x^-}}^{n-1}  \polylog{1-n}{q} \ ,
\end{align}
\end{subequations}
for both $n=2$ and $n=3$. The form of the $\chi\chi^{\dag}$ correlation function is dimension dependent; a brief calculation gives
\begin{align} \nonumber
    \vac{T \{ \chi(x^+,x^-,x) \chi^{\dag}(0) \} } = 
    - & \frac{\sqrt{2} i P_+}{16\pi^2 R (x^-)^2} \bigg( \frac{x^2 }{x^- R} \polylog{-1}{q} + \cot\brac{\frac{x_{\mu} x^{\mu}}{4R \, x^-}} \\
    &+ i \big[ \Theta(x^-) - \Theta(-x^-) \big] \bigg)
\end{align}
when $n=2$, and
\begin{equation}
    \vac{T \{ \chi(x^+,x^-,x) \chi^{\dag}(0) \} } = 
    - \frac{\sqrt{2} P_+}{32 \pi^3 R^2 (x^-)^3} \brac{ \frac{x^2}{x^- R} \polylog{-2}{q} - 4i \, \polylog{-1}{q} }
\end{equation}
when $n=3$. If we decompactify the null direction by taking $R\to\infty$, the large-$R$ expansions
\begin{subequations} \label{eq: large R expansions}
\begin{align}
    \polylog{-s}{q} &=  s! \brac{\frac{2i R\, x^-}{x_{\mu} x^{\mu}}}^{s+1} + O(R^{s}) \ , \\
    \cot\brac{\frac{x_{\mu} x^{\mu}}{4R\, x^-}} &= \frac{4R \, x^-}{x_{\mu} x^{\mu}} + O(1) \ ,
\end{align}
\end{subequations}
for $s\in\bb{N}$ give
\begin{equation}
    \lim_{R\to\infty} \brac{\vac{T\{\psi(x^+,x^-,x) \psi^{\dag}(0)\}}} = \frac{i (n-1)!}{2\pi^n R \brac{x_{\mu} x^{\mu}}^n} \brac{\sqrt{2} x^- P_- + \sqrt{2} x^+ P_+ + \gamma_{0i} x^i} \ .
\end{equation}
We see that we have recovered the normal Minkowski spacetime two-point functions \eqref{eq: 4d normalised minkowski 2 point} when $n=2$ and \eqref{eq: 6d normalised minkowski 2 point} when $n=3$. 

While both the $SU(1,n)$ and DLCQ methods can reproduce the Minkowski spacetime two-point function of the free fermion through the resummation of a Kaluza-Klein tower of lower dimensional two-point functions, there are conceptual differences between the approaches. Most notably, while we had to take the $R\to\infty$ limit to recover the two-point function from the DLCQ correlation functions, no such limit was taken on the $SU(1,n)$ side; the original theory was recovered through a diffeomorphism and Weyl transformation, so the mapping between the two is invertible. We also see that the $SU(1,n)$ mode two-point functions exhibit a spatial fall-off, unlike the DLCQ mode two-point functions. It is interesting to ask whether this leads to any improvements in computing quantities in the quantum theory: however, we leave this question to future work.

\subsection{Hints of a Carrollian Limit}
We have seen that taking $R\to\infty$ with the ratio $\frac{k}{R}$ fixed leads to an action with Schr\"odinger symmetry at each level in the Kaluza-Klein tower. It is therefore natural to ask what happens in the opposite limit, in which we define
\begin{subequations}
\begin{align}
    R &= \omega R_+ \ , \\
    k &= \omega k_+ \ ,
\end{align}
\end{subequations}
and take the limit $\omega\to0$ with $R_+$ and $k_+$ fixed. Let us first consider this for the complex scalar action \eqref{eq: phi action}. First rescaling the field to
\begin{equation} \label{eq: field rescaling}
    \phi_k = \frac{\tilde{\phi}_{k}}{\sqrt{2\pi R}}
\end{equation}
gives (dropping the pluses, tildes, and field subscript to simplify notation)
\begin{align} \nonumber
    S_k^{(\phi)} =  \int d^{2n-1}x \bigg(& \frac{2i k}{R} \bar{\phi} \partial_- \phi - \partial_i \bar{\phi} \partial_i \phi + \inv{2 \omega} \Omega_{ij} x^j \Big(\partial_i \bar{\phi} \partial_- \phi \\
    &+ \partial_- \bar{\phi} \partial_i \phi \Big)
    - \frac{x^2}{4 \omega^2 R^2} \partial_- \bar{\phi} \partial_-\phi \bigg) \ .
\end{align}
We see that there are two sets of terms that diverge; we must deal with both of these to recover a sensible theory in the $\omega \to 0$ limit. As the $O(\omega^{-2})$ term is a square, the most general way to remove the leading divergence is to impose
\begin{equation}
    x^i \partial_- \phi = 0 \ .
\end{equation}
 If we additionally require that our field is smooth we can drop the factor of $x^i$, and we shall assume this is true from here onwards. The constraint also removes the divergences arising from the $O(\omega^{-1})$ terms and imposing it is enough to render the theory finite as $\omega\to0$. We therefore find the action
 \begin{equation} \label{eq: carroll scalar action}
     S_k^{(\phi)} = \int d^{2n-1}x \brac{\bar{H} \partial_- \phi + \partial_- \bar{\phi} H - \partial_i \bar{\phi} \partial_i \phi} 
 \end{equation}
after taking the limit, where $H$ is a Lagrange multiplier field implementing the constraint. This is the action of a 'spacelike'\footnote{Also referred to as 'magnetic' \cite{deBoer:2023fnj}.} scalar field with conformal Carroll spacetime symmetry \cite{Baiguera:2022lsw}.

Let us do the same with the fermion action \eqref{eq: fermion level k action}, where we assume that $k$ is non-zero. If we again rescale our fields as in \eqref{eq: field rescaling}, this becomes
\begin{align} \nonumber
    S_k = i \int d^{2n-1}x \bigg(& \sqrt{2} \lambda^{\dag} \partial_- \lambda + \chi^{\dag} \gamma_{0i} \partial_i \lambda + \lambda^{\dag} \gamma_{0i} \partial_i \chi - \frac{\sqrt{2} i k}{R} \chi^{\dag} \chi \\
    &- \inv{2\omega} \Omega_{ij} x^j \brac{\chi^{\dag} \gamma_{0i} \partial_- \lambda + \lambda^{\dag} \gamma_{0i} \partial_- \chi} \bigg) \ .
\end{align}
In contrast to the scalar case all divergent terms are of the same order in $\omega$. To obtain a finite action we must impose
\begin{equation}
    \Omega_{ij} x^j \brac{\chi^{\dag} \gamma_{0i} \partial_- \lambda + \lambda^{\dag} \gamma_{0i} \partial_- \chi} = 0 \ .
\end{equation}
This is quite a complicated constraint, and we will restrict ourselves to working with the subset of field configurations that satisfy the simpler conditions
\begin{subequations} \label{eq: fermion constraints}
\begin{align}
    \partial_- \lambda &= 0 \ , \\
    \partial_- \chi &= 0 \ .
\end{align}
\end{subequations}
It is valid to ask if there are any interesting solutions missed by imposing these stronger constraints, though we will not pursue this here. Introducing fermionic Lagrange multiplier fields $\rho$ and $\eta$ for our constraints gives the action
\begin{equation}
    S_k = i \int d^{2n-1}x \bigg( \chi^{\dag} \gamma_{0i} \partial_i \lambda + \lambda^{\dag} \gamma_{0i} \partial_i \chi - \frac{\sqrt{2}i k}{R} \chi^{\dag} \chi + \eta^{\dag} \partial_- \lambda + \partial_- \lambda^{\dag} \eta + \rho^{\dag} \partial_- \chi + \partial_- \chi^{\dag} \rho
    \bigg)
\end{equation}
in the $\omega\to0$ limit, from which we see that the equation of motion for $\chi$ is now
\begin{equation}
    \chi = \frac{i R}{\sqrt{2} k} \brac{\partial_- \rho - \gamma_{0i} \partial_i \lambda} \ .
\end{equation}
As this is still algebraic in $\chi$ we can substitute it back into the action to obtain
\begin{equation} \label{eq: carroll fermion 1}
    S_k = \frac{R}{\sqrt{2} k} \int d^{2n-1}x \brac{ \partial_- \rho^{\dag} \partial_- \rho + \eta^{\dag} \partial_- \lambda + \partial_- \lambda^{\dag} \eta - \partial_i \lambda^{\dag} \partial_i \lambda } \ .
\end{equation}
We recognise the terms involving $\lambda$ as those of a 'spacelike' Carrollian field, while the action for $\rho$ is that of a 'timelike'\footnote{Or 'electric'.} Carrollian field. Alternatively, we could instead impose
\begin{equation}
    \chi = 0 \ .
\end{equation}
The action then reduces to the simple form
\begin{equation} \label{eq: carroll fermion 2}
    S_k = \sqrt{2} i \int d^{2n-1}x \, \lambda^{\dag} \partial_- \lambda \ ,
\end{equation}
which was first constructed in \cite{Bagchi:2022eui} as a Carrollian limit of a free massless fermion.

We can understand the emergence of Carrollian physics using the $2n$-dimensional spacetime. Recall that after a Weyl transformation the metric \eqref{eq: metric} is
\begin{equation}
    ds^2 = -2 dx^+ \brac{dx^- + \inv{2} \Omega_{ij} x^j dx^i} + dx^i dx^i \ ,
\end{equation}
with our coordinates taking values in the ranges $x^+\in(-\pi R, \pi R)$ and $x^-,x^i \in (-\infty,\infty)$. If we try and take the limit $R\to0$ we find the term involving $\Omega$ diverges. However, we can remove all $R$-dependence in the metric by working with the coordinates
\begin{subequations} \label{eq: map}
\begin{align}
    v &= \frac{x^+}{R} \ , \\
    u &= R x^- \ ,
\end{align}
\end{subequations}
for which the metric becomes
\begin{equation}
    ds^2 = - 2 dv \brac{du + \inv{2} \Tilde{\Omega}_{ij} x^j dx^i} + dx^i dx^i \ ,
\end{equation}
with $\Tilde{\Omega}_{ij} = R \Omega_{ij}$. The new coordinates have the ranges $v\in(-\pi,\pi)$ and $u\in(-\infty,\infty)$, so all reference to $R$ has disappeared from the theory. Naively, it appears that there should be no issues with taking $R\to0$. The problem comes from the fact that \eqref{eq: map} is not defined when $R=0$. In particular, when $R=0$ the range of $x^+$ collapses to the single point $x^+=0$ whereas the range of $v$ remains finite, so we are mapping between manifolds of different dimension. We can remedy this by restricting ourselves to the hypersurface $v=0$ in the $R$-independent spacetime parameterised by $(v,u,x^i)$. This is a null hypersurface of a $2n$-dimensional Lorentzian spacetime and defines a $(2n-1)$-dimensional Carrollian spacetime \cite{Hartong:2015xda}. A field theory on this background will therefore be invariant under the $(2n-1)$-dimensional Carroll group. We note that none of the details here depend on $\Omega$, and the same structure will appear in more conventional null circle compactifications.

Let us illustrate this discussion with an example. Consider a complex scalar field $\hat{\phi}$. Its Lagrangian on the 'extended' spacetime is
\begin{equation}
    \cal{L} = \partial_v \hat{\bar{\phi}} \partial_u \hat{\phi} + \partial_u \hat{\bar{\phi}} \partial_v \hat{\phi} + \inv{2} \tilde{\Omega}_{ij} x^j \brac{\partial_i \hat{\bar{\phi}} \partial_u \hat{\phi} + \partial_u \hat{\bar{\phi}} \partial_i \hat{\phi}} - \frac{x^2}{4} \partial_u \hat{\bar{\phi}} \partial_u \hat{\phi} - \partial_i \hat{\bar{\phi}} \partial_i \hat{\phi} \ .
\end{equation}
Expanding $\hat{\phi}$ about $v=0$ as
\begin{equation}
    \hat{\phi} = \phi + v H + O(v^2) \ ,
\end{equation}
we can pull $\cal{L}$ back to the $v=0$ hypersurface to obtain the action
\begin{equation}
    S = \int d^{2n-1}x \brac{ \bar{H} \partial_u \phi + \partial_u \bar{\phi} H + \inv{2} \tilde{\Omega}_{ij} x^j \brac{\partial_i \bar{\phi} \partial_u \phi + \partial_u \bar{\phi} \partial_i \phi} - \frac{x^2}{4} \partial_u \bar{\phi} \partial_u \phi - \partial_i \bar{\phi} \partial_i \phi } \ .
\end{equation}
Note that we must treat $H$ and $\phi$ as independent fields. We see that $\bar{H}$ is a Lagrange multiplier that imposes the constraint
\begin{equation}
    \partial_u \phi = 0 \ ,
\end{equation}
so we can rewrite the action in the equivalent form
\begin{equation}
    S = \int d^{2n-1}x \brac{ \bar{H} \partial_u \phi + \partial_u \bar{\phi} H - \partial_i \bar{\phi} \partial_i \phi } \ .
\end{equation}
We recognise this as the action \eqref{eq: carroll scalar action} for a magnetic Carroll scalar field theory. Similarly, performing the same calculation for a fermion recovers the action \eqref{eq: carroll fermion 2}. It does not appear possible to construct the action \eqref{eq: carroll fermion 1} using this approach, as this would require the initial Lagrangian to have terms that are second-order in derivatives to get the correct Lagrange multiplier structure.

The limit $R\to0$ is somewhat less natural than the previously studied $R\to\infty$ limit. For instance, the correlation functions of both the scalar and fermion $SU(1,n)$ theories vanish once it is taken. It would be interesting to see if the limit (possibly modified) can be used to make non-trivial statements about Carrollian field theories using quantities computed in $SU(1,n)$ theories, or if the relationship between the two is limited to the construction of Carrollian actions from their $SU(1,n)$ counterparts.

\section{Reduction of the 2d Fermion}
\label{sec: 2d ferm reduction}

So far we have only been interested in theories in $D=4$ and $D=6$. As there are no known interacting CFTs when $D>6$ there is little point in analysing higher-dimensional cases. This still leaves us with the exceptional case $D=2$, which we examine now.

The coordinate transformation \eqref{eq: coord transformation} behaves differently in 2d as it does in other dimensions since there are no transverse directions, and hence no $\Omega$-deformation. In this case, the transformation is a conformal transformation of Minkowski spacetime and the resulting field theory remains a 2d Lorentzian CFT. The null reduction is then a geometrically-implemented DLCQ and falls somewhere between the methods of sections \ref{sec: corr} and \ref{sec: DLCQ}. We would like to determine to what extent the correlation functions of the 2d theory can be recovered from the reduced 1d theory.

Let us be more precise. We start with the 2d Minkowski metric in lightcone coordinates,
\begin{equation}
    ds^2 = - 2 d\hat{x}^+ d\hat{x}^- \ .
\end{equation}
The coordinate transformation is
\begin{subequations} \label{eq: 2d transformation}
\begin{align}
    \hat{x}^+ &= 2R \tan \brac{\frac{x^+}{2R}} \ , \\
    \hat{x}^- &= x^- \ ,
\end{align}
\end{subequations}
so the metric becomes
\begin{equation}
    ds^2 = \sec^2\brac{\frac{x^+}{2R}} \brac{- 2 dx^+ dx^-} \ .
\end{equation}
As usual, the coordinate $x^+$ takes values in the interval $(-\pi R, \pi R)$; after a Weyl transformation we can then include the endpoints and take $x^+\in[-\pi R, \pi R]$.

Our field theory is taken to be a single Majorana fermion $\hat{\psi}$ with the action
\begin{equation} \label{eq: 2d action}
    S = - \int d \hat{x}^+ d\hat{x}^- \Bar{\hat{\psi}} \gamma^{\mu} \partial_{\mu} \hat{\psi} \ .
\end{equation}
Using $\gamma_* = \gamma_{01}$ and defining the chiral components
\begin{subequations}
\begin{align}
    \inv{2} \brac{\bbm{1} + \gamma_* } \hat{\psi} &= \hat{\chi} \xi_1 \ , \\
    \inv{2} \brac{\bbm{1} - \gamma_* } \hat{\psi} &= \hat{\lambda} \xi_2 \ , 
\end{align}
\end{subequations}
in terms of orthonormal spinors $\{\xi_i\}$\footnote{Note that this the fields $\hat{\lambda}$ and $\hat{\chi}$ are real Grassmann numbers and possess no spinor indices.}, we can rewrite the action as
\begin{equation}
    S = i\int d\hat{x}^+ d \hat{x}^- \Bigg[ \hat{\chi} \partial_{\hat{+}} \hat{\chi} + \hat{\lambda} \partial_{\hat{-}} \hat{\lambda} \Bigg] \ .
\end{equation}
It's then clear that the action is invariant if the fields transform as
\begin{subequations}
\begin{align}
    \hat{\chi}(\hat{x}) &= \chi(x) \ , \\
    \hat{\lambda}(\hat{x}) &= \cos\brac{\frac{x^+}{2R}} \lambda(x) \ ,
\end{align}
\end{subequations}
under \eqref{eq: 2d transformation}. As discussed in section \ref{sec: action 2}, we should fix the decomposition of $\chi$ and $\lambda$ into modes by requiring that $\hat{\chi}$ and $\hat{\lambda}$ admit Fourier series expansions. This gives
\begin{subequations}
\begin{align}
    \chi(x) &= \sum_{l\in\bb{Z}} e^{-\frac{i l x^+}{R}} \chi_{l} \ , \\
    \lambda(x) &= \sum_{k\in \bb{Z} + \inv{2}} e^{- \frac{i k x^+}{R}} \lambda_k \ .
\end{align}
\end{subequations}
In terms of the modes, the action becomes
\begin{equation}
    S = 2 \pi i R \sum_{l=1}^{\infty} \int dx^- \lambda_{-(l - \inv{2})} \partial_- \lambda_{l-\inv{2}} + 2\pi \sum_{l = 1}^{\infty} l \int dx^- \chi_{-l} \chi_l \ .
\end{equation}
We see that, unlike in higher dimensions, one of the chiral components of our field has become non-dynamical upon performing the mode decomposition. This is expected: recall that we are working with the DLCQ of a 2d theory, which is known to freeze the chiral sector along which we perform the reduction (as pointed out in \cite{Balasubramanian:2009bg}). From here onwards we will restrict our attention to $\lambda$.

If we Fourier transform on $x^-$ with the conventions
\begin{equation}
    \lambda_k(x^-) = \int \frac{d\omega}{2\pi} e^{-i \omega x^-} \Tilde{\lambda}_k(\omega) \ ,
\end{equation} 
the action simplifies to
\begin{equation}
    S_{\lambda} = 2\pi R \sum_{l = 1}^{\infty} \int \frac{d\omega}{2\pi} \, \omega \Tilde{\lambda}_{-(l - \inv{2})}(-\omega) \Tilde{\lambda}_{l - \inv{2}}(\omega) \ .
\end{equation}
We can add the source terms
\begin{equation}
    S_{\xi} = \sum_{l = -\infty}^{\infty} \int dx^- \, \xi_{-(l-\inv{2})} \lambda_{l - \inv{2}}
\end{equation}
to the action, and after making the field redefinition
\begin{equation}
    \Tilde{\lambda}_k(\omega) = \Tilde{\eta}_k(\omega) - \frac{\xi_k(\omega)}{\pi R \omega}
\end{equation}
we find
\begin{equation}
    S_{\lambda} + S_{\xi} = \sum_{l = 1}^{\infty} \int \frac{d\omega}{2\pi} \Bigg[ 2\pi R \omega \, \Tilde{\eta}_{-(l-\inv{2})}(-\omega) \Tilde{\eta}_{l-\inv{2}}(\omega) - \inv{2\pi R \omega} \Tilde{\xi}_{-(l-\inv{2})}(-\omega) \Tilde{\xi}_{l-\inv{2}}(\omega) \Bigg] \ .
\end{equation}
The partition function of the theory is then
\begin{equation}
    \cal{Z}[\xi] = \cal{Z}[0] \exp \brac{- \frac{i}{2\pi R} \sum_{l=1}^{\infty} \int \frac{d\omega}{2\pi} \, \inv{\omega} \Tilde{\xi}_{-(l-\inv{2})}(-\omega) \Tilde{\xi}_{l-\inv{2}} (\omega) }  \ .
\end{equation}
We can take derivatives of this and Fourier transform back to obtain the time-ordered two-point functions
\begin{align} \nonumber
    \vac{T\{ \lambda_k(x^-) \lambda_{-k}(0)\}} &= \frac{i}{R} \int \frac{d\omega_1 d\omega_2}{(2\pi)^2} \,e^{-i \omega_1 x^-} \frac{\delta(\omega_1 + \omega_2)}{\omega_2} \\
    &= -\frac{i}{2\pi R} \int \frac{d\omega}{2\pi} \, \frac{e^{-i \omega x^-}}{\omega}
\end{align}
for both positive and negative $k$. As it stands, this integral is ill-defined; we can fix this using a $k$-dependent $i\epsilon$ prescription
\begin{equation}
    \int \frac{d\omega}{2\pi}\, \frac{e^{- i\omega x^-}}{\omega} \to \int \frac{d\omega}{2\pi}\, \frac{e^{-i\omega x^-}}{\omega + \frac{ i k \epsilon}{R}} \ .
\end{equation}
This can then be easily evaluated using the residue theorem, from which we get
\begin{equation}
    \vac{T\{ \lambda_k(x^-) \lambda_{-k}(0)\}} = - \frac{e^{- \frac{ k \epsilon x^-}{R}} \Theta(x^-)}{2\pi R}
\end{equation}
for $k>0$, and
\begin{equation}
    \vac{T\{ \lambda_k(x^-) \lambda_{-k}(0)\}} = \frac{e^{\frac{ k \epsilon x^-}{R}} \Theta(-x^-)}{2\pi R}
\end{equation}
for $k<0$. These are equivalent to the conditions
\begin{equation}
    \lambda_k(x^-) \ket{\Omega} = 0 \; \; \forall \ k>0 
\end{equation}
and the correlation functions
\begin{equation}
    \vac{\lambda_k(x^-) \lambda_{-k}(0)} = - \frac{e^{- \frac{k \epsilon \abs{x^-}}{R}}}{2\pi R}
\end{equation}
for positive $k$.

Let us use this to reconstruct the two-point function of $\hat{\lambda}$. We have
\begin{align} \nonumber
    \vac{T\{ \hat{\lambda}(\hat{x}^+, \hat{x}^-) \hat{\lambda}(0)\}} = &\cos\brac{\frac{x^+}{2R}} \sum_{l=0}^{\infty} \Bigg[ e^{- \frac{i x^+}{R}\brac{l + \inv{2}}} \vac{T\{ \lambda_{l+\inv{2}}(x^-) \lambda_{-(l+\inv{2})}(0)\}} \\
    &+  e^{ \frac{i x^+}{R}\brac{l + \inv{2}}} \vac{T\{ \lambda_{-(l+\inv{2})}(x^-) \lambda_{l+\inv{2}}(0)\}} 
    \Bigg] \ ,
\end{align}
so substituting in the mode two-point functions gives
\begin{align} \nonumber
    \vac{T\{ \hat{\lambda}(\hat{x}^+, \hat{x}^-) \hat{\lambda}(0)\}} &= \frac{\cos \brac{\frac{x^+}{2R}}}{2\pi R} \sum_{l=0}^{\infty} \Bigg[ - \Theta(x^-) \sqrt{q} q^l + \Theta(-x^-) \frac{q^{-l}}{\sqrt{q}} \Bigg] \\
    &= - \frac{\cos \brac{\frac{x^+}{2R}}}{2\pi R} \frac{\sqrt{q}}{1-q} \ ,
\end{align}
where we've defined
\begin{equation}
    q = e^{- \frac{i}{R}\brac{x^+ - i x^- \epsilon}} \ .
\end{equation}
We note in passing that the $\epsilon$ shift of the exponent ensures convergence of the sum. After some rearrangement, we recover the result (suppressing the $\epsilon$ contribution)
\begin{equation}
    \vac{T\{ \hat{\lambda}(\hat{x}^+, \hat{x}^-) \hat{\lambda}(0)\}} = \frac{i}{2\pi \hat{x}^+}
\end{equation}
for the two-point function of a left-moving free fermion arising from the action \eqref{eq: 2d action}. While we have managed to reconstruct the left-moving sector of the theory, it is impossible to reconstruct the right-moving sector in this way. We should contrast this with the higher-dimensional cases previously discussed, where it was found that the full two-point function can be reconstructed from the null reduction.

\section{\texorpdfstring{$\frak{su}(1,n)$}{su(1,n)} Ward Identities}

\label{sec: WI}

While we have discussed both the fermion action and its associated two-point functions in some depth, we have not yet shown that the theories really are invariant under the spacetime symmetry algebra $\frak{su}(1,n)$ as we have claimed multiple times. We will now remedy this by showing that the two-point functions satisfy the corresponding Ward identities. 

Let us briefly review the $\frak{su}(1,n)$ spacetime symmetry algebra and its associated Ward identities before showing that they are satisfied in our theory. First, consider the 3d theory. The algebra $\frak{su}(1,2)$ is generated by the basis $\{P_-, P_i, B, T, G_i, K\}$ and can be centrally extended by the element $P_+$. Physically, $P_-$ and $P_i$ form a non-Abelian algebra of translations, $B$ is the generator of spatial rotations, $T$ generates Lifshitz scalings, and $G_i$ and $K$ generate non-Lorentzian special conformal transformations. We identify $P_+$ as the generator of translations in the compactified direction in the original theory.
The 5d theory is almost identical; the only addition necessary to obtain $\frak{su}(1,3)$ is an $\frak{su}(2)$ algebra with generators $\{J^{\alpha}\}$. These commute with $B$ and generate the remaining spatial rotations. The Lie brackets between the generators are given in \cite{Lambert:2021nol}, though we will instead work with anti-Hermitian operators for convenience.

To simplify our discussion we will mostly work with primary operators $\{\cal{O}_k\}$ that obey the conditions
\begin{align}
    [G_i, \cal{O}_k(0)] = [K, \cal{O}_k(0)] = 0 \ .
\end{align}
Borrowing from the nomenclature of Lorentzian CFTs, all operators that do not satisfy these conditions are referred to as descendants. The non-derivative part of the action of the algebra on a primary operator at the origin is then
\begin{subequations} \label{eq: rep def}
\begin{align} 
    [ \cal{O}(0), T] &= \Delta \cal{O}(0) \ , \\
    [ \cal{O}(0), P_+] &= -ip \cal{O}(0) \ , \\
    [ \cal{O}(0), B] &= r_{\cal{O}}[B] \cal{O}(0) \ , \\ 
    [ \cal{O}(0), J^{\alpha}] &= r_{\cal{O}}[J^{\alpha}] \cal{O}(0) \ ,
\end{align}
\end{subequations}
where we have assumed $\{r_{\cal{O}}[J^{\alpha}]\}$ form a finite-dimensional irreducible representation of $\frak{su}(2)$, and $r_{\cal{O}}[B]$ is a representation of $\frak{u}(1)$. We will take $\Delta,k\in\bb{C}$.

The Ward identities for the two-point function of primary operator $\cal{O}_1$ and $\cal{O}_{2}$ are then \cite{Lambert:2020zdc}
\begin{subequations}
\begin{align}
    0 =& \brac{p_1 + p_2} \vac{\cal{O}_1(x_1) \cal{O}_2(x_2)}  \ , \\
    0 =& \sum_{a=1}^2 \brac{\frac{\partial}{\partial x^-_a}} \vac{\cal{O}_1(x_1) \cal{O}_2(x_2)} \ , \\
    0 = &\sum_{a=1}^2 \brac{\frac{\partial}{\partial x^i_a} + \inv{2} \Omega_{ij} x_a^j \frac{\partial}{\partial x_a^-}} \vac{\cal{O}_1(x_1) \cal{O}_2(x_2)}  \ , \\
    0 =& \sum_{a=1}^2 \brac{2 x_a^-\frac{\partial}{\partial x^-_a} + x_a^i \frac{\partial}{\partial x_a^i} + \Delta_a} \vac{\cal{O}_1(x_1) \cal{O}_2(x_2)}  \ , \\
    0 = & \sum_{a=1}^2 \brac{-R\Omega_{ij} x_a^i \frac{\partial}{\partial x_a^j} + r_a[B]} \vac{\cal{O}_1(x_1) \cal{O}_2(x_2)}  \ , \\
    0 = & \sum_{a=1}^2 \brac{\Bar{\eta}^{\alpha}_{ij} x_a^i \frac{\partial}{\partial x_a^j} + r_a[J^{\alpha}]} \vac{\cal{O}_1(x_1) \cal{O}_2(x_2)} \ , \\ \nonumber
    0 = & \sum_{a=1}^2 \Bigg( \brac{\inv{2} \Omega_{ij} x_a^- x_a^j - \inv{8R^2}x_a^2 x_a^i} \frac{\partial}{\partial x_a^-} + x_a^- \frac{\partial}{\partial x_a^i} + \inv{4} \bigg( 2 \Omega_{ik} x_a^k x_a^j \\ \nonumber
    &+ 2 \Omega_{jk} x_a^k x_a^i - \Omega_{ij} x_a^2 \bigg) \frac{\partial}{\partial x_a^j} + \inv{2} \Delta_a \Omega_{ij} x_a^j - i p_a x_a^i  \\ \label{eq: Gi WI}
    &+  \brac{\frac{n+1}{n-1}} \frac{x_a^i r_a[B]}{2R}  - \inv{2} \Omega_{ik} \bar{\eta}^{\alpha}_{jk} x_a^j r_a[J^{\alpha}] \Bigg) \vac{\cal{O}_1(x_1) \cal{O}_2(x_2)} \ ,  \\ \nonumber
    0 = & \sum_{a=1}^2 \Bigg( \brac{2(x_a^-)^2 - \frac{x_a^4}{8R^2}} \frac{\partial}{\partial x_a^-} + \brac{ \inv{2} \Omega_{ij} x_a^j x_a^2 + 2x_a^- x_a^i } \frac{\partial}{\partial x_a^i} + 2 \Delta_a x_a^- \\ \label{eq: K WI}
    &- i p_a x_a^2 + \brac{\frac{n+1}{n-1}}\frac{x_a^2 r_a[B]}{2R}  - \inv{2} x_a^i x_a^j \Omega_{ik} \bar{\eta}^{\alpha}_{jk} r_a[J^{\alpha}] \Bigg) \vac{\cal{O}_1(x_1) \cal{O}_2(x_2)} \ ,
\end{align}
\end{subequations}
where any terms with either $r_a[J^{\alpha}]$ or $\bar{\eta}_{ij}^{\alpha}$ are ignored for the $n=2$ identities. Note that the condition that the operators are primary has only been used in the final two identities. 

It will be convenient to introduce the complex coordinate
\begin{equation}
    z_{12} = x^-_1 - x_2^- + \inv{2}\Omega_{ij} x_1^i x_2^j + \frac{i}{4 R} (x_1 - x_2)^2 \ ,
\end{equation}
which is the translated version of \eqref{eq: z def}. We then immediately have
\begin{align}
    \sum_{a=1}^2 \brac{\frac{\partial}{\partial x_a^-}} f(z_{12}, \Bar{z}_{12}) &= 0 \ , \\
    \sum_{a=1}^2 \brac{\frac{\partial}{\partial x^i_a} + \inv{2} \Omega_{ij} x_a^j \frac{\partial}{\partial x_a^-}} f(z_{12}, \Bar{z}_{12}) &= 0 \ , \\
    \sum_{a=1}^2 \brac{ - R \Omega_{ij} x_a^i \frac{\partial}{\partial x_a^j} } f(z_{12}, \Bar{z}_{12}) &= 0 \ , \\
    \sum_{a=1}^2 \brac{ \Bar{\eta}^{\alpha}_{ij} x_a^i \frac{\partial}{\partial x_a^j} } f(z_{12}, \Bar{z}_{12}) &= 0 \ ,
\end{align}
for any function $f$. Using the notation $\cal{L}_i$ and $\cal{L}$ for the differential operators in \eqref{eq: Gi WI} and \eqref{eq: K WI}, similar computations give
\begin{align}
    \cal{L}_i f(z_{12}, \Bar{z}_{12}) &= \frac{i}{2R} (x_1 - x_2)^i \brac{z_{12} \partial - \Bar{z}_{12} \Bar{\partial}}f + \inv{2}\Omega_{ij} (x_1 + x_2)^j \brac{z_{12} \partial + \Bar{z}_{12} \Bar{\partial}} f \ , \\
    \cal{L} f(z_{12} , \Bar{z}_{12}) &= 2(x_1^- + x_2^-) \brac{z_{12} \partial + \Bar{z}_{12} \Bar{\partial}} f + \frac{i}{2R} \brac{x_1^2 - x_2^2} \brac{z_{12} \partial - \Bar{z}_{12} \Bar{\partial}}f \ .
\end{align}

\subsection{\texorpdfstring{$\frak{su}(1,2)$}{su(1,2)} Invariance in 3d}

Let us compute the constraints imposed by the Ward identities on the correlation functions of the 3d theory and check that they possess $\frak{su}(1,2)$ spacetime symmetry. We will focus on $k>0$, but the extension to $k<0$ is obvious; we shall comment on the $k=0$ case later. In this regime, the equation of motion \eqref{eq: chi eom non-zero k} for $\chi_k$ allows it to be expressed in terms of derivatives of $\lambda_k$. In other words, $\chi_k$ is a descendant field. This means it is sufficient to focus on the two-point function of the primary field $\lambda_k$. The field $\lambda_k$ only has a single complex degree of freedom, so we can take $r_{\lambda_k}[B]$ to be an irreducible representation of $\frak{u}(1)$.

Assuming translational invariance of our theory (which is simple to check from the action), the two-point function we determined previously can be written as
\begin{equation}
    G^{\lambda \lambda}_k(x_1^-, x_1; x_2^-,x_2) = - \frac{ik}{4\sqrt{2} \pi^2 R^2} z_{12}^{k-1} \bar{z}_{12}^{-k} P_- \ .
\end{equation}
The non-trivial Ward identities are then
\begin{subequations}
\begin{align}
    0 = & \ p_{\lambda_k} + p_{\lambda^{\dag}_k} \ , \\
    2 = & \ \Delta_{\lambda_k} + \Delta_{\lambda^{\dag}_k} \ , \\
    0 = & \ r_{\lambda_k}[B] + r_{\lambda_k^{\dag}}[B] \ , \\ \nonumber
    0 = & \  \frac{i}{R}  \brac{k - \inv{2} - R p_{\lambda_k} - \frac{3i}{2} r_{\lambda_k}[B] } \brac{  x_1 - x_2 }^i
    \\
    & \ - \inv{2} (1 - \Delta_{\lambda_k}) \Omega_{ij} \brac{  x_1 - x_2}^j \ , \\ \nonumber
    0 = & \ \frac{i}{R}  \brac{k - \inv{2} - R p_{\lambda_k} - \frac{3i}{2} r_{\lambda_k}[B] } \brac{  x_1^2 - x_2^2} 
    \\
    & \ - 2 (1 - \Delta_{\lambda_k}) \brac{x_1^- - x_2^-} \ ,
\end{align}
\end{subequations}
where we have used the first three identities to simplify the final two.

If we take the Hermitian conjugate of \eqref{eq: rep def} we see that we must have
\begin{subequations}
\begin{align}
    \Delta_{\lambda_k^{\dag}} &= \brac{\Delta_{\lambda_k}}^* \ , \\
    p_{\lambda_k^{\dag}} &= - \brac{p_{\lambda_k}}^* \ , \\
    r_{\lambda_k^{\dag}}[B] &= \brac{r_{\lambda_k}[B]}^* \ .
\end{align}
\end{subequations}
The first Ward identity is then just the statement that $p_{\lambda_k}$ is real, and the third reduces to
\begin{equation}
    r_{\lambda_k}[B] = i \alpha_{\lambda_k} 
\end{equation}
for some real $\alpha_{\lambda_k}$. This means that our charges need to satisfy
\begin{subequations}
\begin{align}
    \Delta_{\lambda_k} &= 1 \ , \\ \label{eq: p alpha relation}
    k - \inv{2} &= R p_{\lambda_k} - \frac{3}{2} \alpha_{\lambda_k} \ .
\end{align}
\end{subequations}
While the requirements of $\frak{su}(1,2)$ symmetry completely fix the value of $\Delta$, this is not true for $p$ and $\alpha$. We can see this from the algebra: since $P_+$ is central, we have the one-parameter family of reparameterisations
\begin{subequations}
\begin{align}
    B' &= B + c P_+ \ , \\
    P_+' &= P_+ + \frac{3c}{2R} P_+ \ ,
\end{align}
\end{subequations}
that leave the algebra invariant. This corresponds to the shift
\begin{subequations}
\begin{align}
    \alpha_{\lambda_k}' &= \alpha_{\lambda_k} + c p_{\lambda_k} \ , \\
    p_{\lambda_k}' &= p_{\lambda_k} + \frac{3 c}{2R} p_{\lambda_k} \ ,
\end{align}
\end{subequations}
in the charges. It's easy to see that these transformations leave \eqref{eq: p alpha relation} invariant. We are therefore allowed to choose $p_{\lambda_k}$ to be any (real) convenient value. It is natural to take
\begin{equation} \label{eq: p param}
    p_{\lambda_k} = \frac{k + \delta}{R} \ ,
\end{equation}
where $\delta$ is a real number that is independent of $k$; with this choice the representation of $B$ is the same for all $\lambda_k$, giving
\begin{equation}
    \alpha_{\lambda_k} = \frac{2\delta + 1}{3} \ .
\end{equation}

Since all Ward identities except \eqref{eq: K WI} and $\eqref{eq: Gi WI}$ are also valid for non-primary operators, they should still hold for our other two-point functions. Let us consider this for
\begin{equation}
    G^{\lambda \chi}_k(x_1^-, x_1; x_2^-,x_2) = -\frac{ik}{8\pi^2 R^2} \gamma_{0i} (x_1 - x_2)^i z_{12}^{k-1} \bar{z}_{12}^{-(k+1)}  P_+ \ .
\end{equation}
The non-trivial identities are
\begin{subequations}
\begin{align}
    p_{\lambda_k} + p_{\chi_k^{\dag}} &= 0 \ , \\
    \Delta_{\lambda_k} + \Delta_{\chi_k^{\dag}} &= 3 \ , \\
    r_{\lambda_k}[B] + r_{\chi_k^{\dag}}[B] &= i \ .
\end{align}
\end{subequations}
Using the results for $\lambda_k$, we find
\begin{subequations}
\begin{align}
    p_{\chi_k} &= p_{\lambda_k} \ , \\ 
    \Delta_{\chi_k} &= 2 \ , \\
    r_{\chi_k}[B] &=  \frac{2i(\delta - 1)}{3} \ .
\end{align}
\end{subequations}
The computation for the other two-point functions proceeds similarly, and one finds that the identities are satisfied with these charges.

The situation is slightly different when $k=0$. The equation of motion for $\chi_0$ does not relate it to $\lambda_0$, so at this level both $\lambda_0$ and $\chi_0$ are primary fields. We should therefore check that all Ward identities are satisfied for
\begin{equation}
    G^{\chi \chi}_0(x_1^-, x_1; x_2^-,x_2) = - \frac{P_+}{4\sqrt{2} \pi^2 R \, \bar{z}_{12}^{2}} \ .
\end{equation}
Taking $p_{\chi_0}$ and $ \Delta_{\chi_0}$ to be real, the non-trivial conditions are
\begin{subequations}
\begin{align}
   \Delta_{\chi_0} &= 2 \ , \\
   R p_{\chi_0} + \frac{3i}{2} r_{\chi_0}[B] &= 1 \ .
\end{align}
\end{subequations}
If we ask that $p_{\chi_0}$ is consistent with our parameterisation \eqref{eq: p param} for $k=0$ the second condition is
\begin{equation}
    r_{\chi_0}[B] = \frac{2i(\delta - 1)}{3} \ ,
\end{equation}
so the charges found for $k\neq0$ are consistent with $\chi_0$'s status as a primary field.

It is interesting to ask which value of $\delta$ naturally occurs when we obtain the theory through the reduction of a 4d Lorentzian theory. As $B$ descends from the spatial rotations of the 4d theory, which are generated by the matrices $\{\gamma_{ij}\}$, the only possible tensorial structure is
\begin{align} \nonumber
    r_{\psi_k}[B] &= - \frac{R \beta_{\psi_k}}{2} \Omega_{ij} \gamma_{ij} \\
    & = i \beta_{\psi_k} \gamma_{01} \ ,
\end{align}
where anti-Hermiticity of $r_{\psi_k}[B]$ constrains $\beta_{\psi_k}$ to be real. The charges of $\lambda_k$ and $\chi_k$ under the rotation must then differ by a sign, which forces us to take
\begin{equation}
    \delta = \inv{4}
\end{equation}
and hence
\begin{align}
    p_{\lambda_k} &= \inv{R} \brac{k + \inv{4}} \ , \\
    \alpha_{\lambda_k} &= \inv{2} \ .
\end{align}

\subsection{\texorpdfstring{$\frak{su}(1,3)$}{su(1,3)} Invariance in 5d}

We will now perform the same analysis for the 5d theory. As the correlation functions vanish when $k=0$ we only need to focus on the $k>0$ modes, meaning the only primary fields are the $\lambda_k$ with translated correlation functions
\begin{align} \nonumber
    G^{\lambda \lambda}_k(x_1^-, x_1; x_2^-,x_2) &= - \frac{k}{8 \sqrt{2} \pi^3 R^3} P_- \brac{(k+1) z_{12}^k \bar{z}_{12}^{-(k+2)} P_-^{\Gamma} + (k-1) z_{12}^{k-2} \bar{z}_{12}^{-k} P_+^{\Gamma} } \\
    & \equiv P_- \brac{\cal{A} z_{12}^k \bar{z}_{12}^{-(k+2)} P_-^{\Gamma} - \cal{B} z_{12}^{k-2} \bar{z}_{12}^{-k} P_+^{\Gamma} }\ .
\end{align}
Our intuition from the calculation in the 3d case leads us to the assumption that $p_{\lambda_k}$ and $\Delta_{\lambda_k}$ are real. This will turn out to be correct, and we will assume it from here onwards for convenience. A crucial difference from the 3d case is that $\lambda_k$ has more than one degree of freedom, so we can no longer assume $r_{\lambda_k}[B]$ is irreducible.

The non-trivial constraints imposed by the $\frak{su}(1,3)$ Ward identities are
\begin{subequations}
\begin{align}
    2 = & \ \Delta_{\lambda_k} \ , \\
    0 = & \ r_{\lambda_k}[B] P_- P_-^{\Gamma} +  P_- P_-^{\Gamma} r^{\dag}_{\lambda_k}[B] = r_{\lambda_k}[B] P_- P_+^{\Gamma} +  P_- P_+^{\Gamma} r^{\dag}_{\lambda_k}[B] \ , \\
    0 = & \ r_{\lambda_k}[J^{\alpha}] P_- P_-^{\Gamma} +  P_- P_-^{\Gamma} r^{\dag}_{\lambda_k}[J^{\alpha}] =  r_{\lambda_k}[J^{\alpha}] P_- P_+^{\Gamma} +  P_- P_+^{\Gamma} r^{\dag}_{\lambda_k}[J^{\alpha}] \ , \\ \nonumber
    0 = & \  \frac{i \brac{k - R p_{\lambda_k}} (x_1 - x_2)^i }{R} G_{k}^{\lambda \lambda}  + \brac{\frac{\delta_{ij}}{R} r_{\lambda_k}[B] - \inv{2} \Omega_{ik} \bar{\eta}^{\alpha}_{jk} r_{\lambda_k}[J^{\alpha}]} x_1^j G_k^{\lambda \lambda} \\ \nonumber
    & \ + x_2^j G_k^{\lambda \lambda} \brac{\frac{\delta_{ij}}{R} r^{\dag}_{\lambda_k}[B] - \inv{2} \Omega_{ik} \bar{\eta}^{\alpha}_{jk} r^{\dag}_{\lambda_k}[J^{\alpha}]}
     + \frac{i(x_1-x_2)^i P_-}{R} 
    \bigg( \cal{A} z_{12}^k \bar{z}_{12}^{-(k+2)} P_-^{\Gamma} \\
    & \ - \cal{B} z_{12}^{k-2} \bar{z}_{12}^{-k} P_+^{\Gamma} \bigg)
    \ , \\ \nonumber
    0 = & \  \frac{i \brac{k - R p_{\lambda_k}} (x_1^2 - x_2^2) }{R}  G_{k}^{\lambda \lambda}  + \brac{\frac{\delta_{ij}}{R} r_{\lambda_k}[B] - \inv{2} \Omega_{ik} \bar{\eta}^{\alpha}_{jk} r_{\lambda_k}[J^{\alpha}]} x_1^i x_1^j G_k^{\lambda \lambda} \\ \nonumber
    & \ + x_2^i x_2^j G_k^{\lambda \lambda} \brac{\frac{\delta_{ij}}{R} r^{\dag}_{\lambda_k}[B] - \inv{2} \Omega_{ik} \bar{\eta}^{\alpha}_{jk} r^{\dag}_{\lambda_k}[J^{\alpha}]} + \frac{i(x_1^2-x_2^2) P_-}{R} 
    \bigg( \cal{A} z_{12}^k \bar{z}_{12}^{-(k+2)} P_-^{\Gamma} \\ 
    & \ - \cal{B} z_{12}^{k-2} \bar{z}_{12}^{-k} P_+^{\Gamma} \bigg)
    \ ,
\end{align}
\end{subequations}
where we have used the first equality to simplify the final two. The natural tensor structures to put forward for our representations are
\begin{subequations}
\begin{align}
    r_{\lambda_k}[B] &= i \alpha_k \Gamma \ , \\
    r_{\lambda_k}[J^{\alpha}] &= \beta_k \bar{\eta}^{\alpha}_{ij} \gamma_{ij} \ ,
\end{align}
\end{subequations}
up to some constants $\alpha_k$ and $\beta_k$. Since
$\bar{\eta}^{\alpha}_{ij} \gamma_{ij}$ and $\Gamma$ commute, as can be checked using the properties of contractions between self-dual and anti-self-dual 't Hooft symbols \cite{tHooft:1976snw}, the second and third Ward identities are automatically satisfied if both $\alpha_k$ and $\beta_k$ are real. However, we have the additional requirement that $\{r_{\lambda_{k}}[J^{\alpha}]\}$ forms a representation of the $\frak{su}(2)$ algebra
\begin{equation}
    [J^{\alpha}, J^{\beta}] = - 2 \epsilon^{\alpha\beta\gamma} J^{\gamma} \ .
\end{equation}
Since
\begin{equation}
    \big[r_{\lambda_k}[J^{\alpha}], r_{\lambda_k}[J^{\beta}] \big] = - 8 \beta^2 \epsilon^{\alpha\beta\gamma} r_{\lambda_k}[J^{\gamma}] \ ,
\end{equation}
we see that this is only true if
\begin{equation}
    r_{\lambda_k}[J^{\alpha}] = \inv{4} \Bar{\eta}^{\alpha}_{ij} \gamma_{ij} \ .
\end{equation}

Using the identities
\begin{subequations}
\begin{align}
    \bar{\eta}^{\alpha}_{ij} \bar{\eta}^{\alpha}_{kl} &= \delta_{ik} \delta_{jl} - \delta_{il} \delta_{jk} - \epsilon_{ijkl} \ , \\
    \gamma_{ijkl} &= - \epsilon_{ijkl} \gamma_{01} \gamma_* \ ,
\end{align}
\end{subequations}
a short computation shows
\begin{equation}
    \Omega_{ik} \bar{\eta}^{\alpha}_{jk} r_{\lambda_k}[J^{\alpha}] P_- = \Omega_{ik} \bar{\eta}^{\alpha}_{jk}  P_- r^{\dag}_{\lambda_k}[J^{\alpha}] = 0 \ ,
\end{equation}
so the terms involving $r_{\lambda_k}[J^{\alpha}]$ in the final two Ward identities vanish. As both have a near-identical structure, let us focus on the first: the remaining terms evaluate to
\begin{equation}
    \frac{i \brac{k - p_{\lambda_k}} (x_1 - x_2)^i }{R} G_{k}^{\lambda \lambda} + \frac{i(x_1-x_2)^i P_-}{R} \brac{1-\alpha_{\lambda_k}}
    \bigg( \cal{A} z_{12}^k \bar{z}_{12}^{-(k+2)} P_-^{\Gamma} - \cal{B} z_{12}^{k-2} \bar{z}_{12}^{-k} P_+^{\Gamma} \bigg) = 0 \ ,
\end{equation}
and we must therefore take
\begin{subequations}
\begin{align}
    p_{\lambda_k} &= \frac{k}{R} \ , \\
    \alpha_{\lambda_k} &= 1 \ .
\end{align}
\end{subequations}

As with the 3d case, the $\frak{u}(1)$ charges of our theory are not uniquely defined by the Ward identities. We can add any multiple of the identity to $r_{\lambda_k}[B]$ in the form
\begin{equation}
    r_{\lambda_k}[B] = i \brac{ \Gamma + c \bbm{1} }
\end{equation}
with $c\in\bb{R}$ without spoiling any of its required properties; the only change is to shift the value of $p_{\lambda_k}$ to
\begin{equation}
    p_{\lambda_k} = \frac{k + c}{R} \ .
\end{equation}
The discussion about the rotation generator inherited from the higher dimensional theory in the 3d theory also applies here; if we require that $r_{\lambda_k}[B]$ is a linear combination of the spinorial rotation generators $\{\gamma_{ij}\}$ then we must impose $c=0$, which we will do for the remainder of this section.

Let us use the Ward identities valid for non-primary two-point functions to fix the representations of $\chi_k$. Using the correlation function
\begin{align} \nonumber
    G^{\lambda \chi}_k(x_1^-, x_1; x_2^-,x_2) &= \frac{ik}{32\pi^3 R^3} \gamma_{0i} P_+ z_{12}^{k-2} \Bar{z}_{12}^{-(k+2)} \Big( z_{12}(1+k) \brac{i \delta_{ij} + R \Omega_{ij}} \\ \nonumber
    & \hspace{1.0cm}- \Bar{z}_{12} (1-k) \brac{i \delta_{ij} - R \Omega_{ij}} \Big) (x_1 - x_2)^j \\
    &\equiv \gamma_{0i} P_+ \Big( \cal{A}\brac{i \delta_{ij} + R\Omega_{ij}} - \cal{B} \brac{i \delta_{ij} - R\Omega_{ij}} \Big) (x_1 - x_2)^j \ ,
\end{align}
we get the constraints
\begin{subequations}
\begin{align}
    5 =& \ \Delta_{\lambda_k} + \Delta_{\chi_k} \ , \\ \nonumber
    0 =& \ \gamma_{0i} P_+ R \Omega_{ij} \Big( \cal{A}\brac{i \delta_{jk} + R\Omega_{jk}} - \cal{B} \brac{i \delta_{jk} - R\Omega_{jk}} \Big) (x_1 - x_2)^k \\
    & \ + r_{\lambda_k}[B] G^{\lambda \chi}_k + G^{\lambda\chi}_k r_{\chi_k}^{\dag}[B] \ , \\
    0 =& \ -\gamma_{0i} P_+ \Bar{\eta}^{\alpha}_{ij} \Big( \cal{A}\brac{i \delta_{jk} + R\Omega_{jk}} - \cal{B} \brac{i \delta_{jk} - R\Omega_{jk}} \Big) (x_1 - x_2)^k \\
    & \ + r_{\lambda_k}[J^{\alpha}] G^{\lambda \chi}_k + G^{\lambda\chi}_k r_{\chi_k}^{\dag}[J^{\alpha}] \ .
\end{align}
\end{subequations}
It's then easy to check that these are satisfied if
\begin{subequations}
\begin{align}
    \Delta_{\chi_k} &= 3 \ , \\
    r_{\chi_k}[B] &= i \Gamma \ , \\
    r_{\chi_k}[J^{\alpha}] &= \inv{4} \Bar{\eta}^{\alpha}_{ij} \gamma_{ij} \ .
\end{align}    
\end{subequations}
Since the rotation generators are the same as those for $\lambda_k$ they automatically form representation of $\frak{u}(1)\oplus\frak{su}(2)$. It is straightforward to check that the valid Ward identities for the other two-point functions are consistent with these assignments.

\section{Conclusion}
\label{sec: conclusion}

In this paper we have analysed fermions in $2n-1$ dimensions with $SU(1,n)$ spacetime symmetry. These were constructed through the null reduction of the $2n$-dimensional free fermion CFT on an $\Omega$-deformed Minkowski background. The two-point functions of the theory for $n=2$ and $n=3$ were found using the $(2n-1)$-dimensional remnants of the supersymmetry of a free complex scalar and free fermion in $2n$ dimensions, and the known two-point functions for a scalar field with $SU(1,n)$ symmetry. The full non-compact $2n$-dimensional two-point function was then reconstructed from the Kaluza-Klein tower of $(2n-1)$-dimensional fields. The analogous transformation for a 2d fermion was also considered, and it was found we could only reconstruct the left-moving fermion's correlation function. We examined the $R\to\infty$ DLCQ limit of the $(2n-1)$-dimensional two-point functions, showing that the $2n$-dimensional two-point function could be recovered through resummation, and commented on the $R\to 0$ limit, which gave a theory with Carroll symmetry. Finally, it was shown that that the Ward identities for $SU(1,n)$ spacetime symmetry are satisfied at each level in the Kaluza-Klein tower. 

Though we have only discussed free field theories, it is interesting to compare the way in which $2n$-dimensional supersymmetry arises in the reduced theories with the supersymmetric structure of the 5d $SU(1,3)$-invariant interacting field theories found in \cite{Lambert:2019jwi, Lambert:2020jjm}. For example, if we truncate the action of the 3d theory to the terms involving the fields $(\lambda_k, \chi_k, \phi_{k+\inv{2}})$ then our theory is invariant under the $\xi$-type supercharges, while the $\eta$-type supercharges of the full theory disappear. Similarly, if we only include the fields $(\lambda_k, \chi_k, \phi_{k-\inv{2}})$ this conclusion is reversed. The same story is true in 5d, where retaining only fields with the same value of $k$ preserves the $\xi$-type supercharges and mixing levels preserves the $\eta$-type supercharges. As discussed in the introduction, this is exactly what occurs in the interacting theories; the Lagrangians only include a finite number of fields and are invariant under half the supercharges of the 6d theories they are conjectured to describe. The fact that the full higher-dimensional supercharge is recovered in our free theory lends weight to the proposal that the same happens when the interacting theories are treated non-perturbatively, as must be the case for the theories to provide complete descriptions of 6d SCFTs. We leave it to future work to show that the full superconformal symmetries of the free theory match the structure of the 5d interacting theories, i.e. that the conformal supercharges are present at each level in the reduced theory.

So far, the aforementioned 5d theories are the only known interacting field theories with $SU(1,n)$ spacetime symmetry for $n\geq 2$. It would be interesting to find other examples; the methods outlined in this work and in \cite{Lambert:2021nol} allow any 4d Lagrangian CFT to be reduced on the $\Omega$-deformed background, yielding a 3d theory with an infinite number of fields and $SU(1,2)$ spacetime symmetry at each level. The task would then be to try and find a non-trivial truncation of the theory to a finite number of fields. The obvious candidate 4d theory is $\cal{N}=4$ SYM, which could help elucidate the field-theoretic formulation of Spin Matrix Theories with $PSU(1,2|3)$ symmetry \cite{Baiguera:2022pll}. A larger class of theories that could also be studied are Lagrangian 4d $\cal{N}=2$ SCFTs, which were classified in \cite{Bhardwaj:2013qia}. Finding the structure of the reduced theories associated with these could shed light on applications of $SU(1,2)$ theories to non-Lagrangian 4d $\cal{N}=2$ SCFTs, such as the class $\cal{S}$ theories \cite{Gaiotto:2009we, Gaiotto:2009hg}.

\acknowledgments
We thank Neil Lambert for the suggestion of this topic and many helpful discussions. J.S. is supported by the STFC studentship ST/W507556/1.

\appendix
\section{Fermion Conventions}
\label{sec: conv}
We will follow the spinor conventions of \cite{Freedman:2012zz} throughout. In particular, we define the Dirac conjugate of a spinor field by
\begin{equation}
    \Bar{\psi} = i \psi^{\dag} \gamma^0 \ ,
\end{equation}
the covariant derivative of a spinor field by
\begin{equation}
    \nabla_{\mu} \psi = \brac{\partial_{\mu} + \inv{4} \omega_{\mu}^{ab} \gamma_{ab}} \ ,
\end{equation}
and the chiral matrix $\gamma_*$ by
\begin{equation}
    \gamma_* = (-i)^{n-1} \gamma_0 \gamma_1 ... \gamma_{n-1} \ .
\end{equation}
When we need an explicit basis for the gamma matrices, we shall use
\begin{subequations}
\begin{align}
    \gamma^{(4d)}_0 &= \begin{pmatrix}
        0 & \bbm{1}_2 \\
        - \bbm{1}_2 & 0
    \end{pmatrix} \ , \\
    \gamma^{(4d)}_i &= \begin{pmatrix}
        0 & \sigma_i \\
        \sigma_i & 0 
    \end{pmatrix} \ ,
\end{align}
\end{subequations}
in 4d, where $\{\sigma_i\}$ are the Pauli matrices, and
\begin{subequations}
\begin{align}
    \gamma^{(6d)}_0 &= \begin{pmatrix}
        0 & 0 & 0 & \bbm{1}_2 \\
        0 & 0 & \bbm{1}_2 & 0 \\
        0 & - \bbm{1}_2 & 0 & 0 \\
        -\bbm{1}_2 & 0 & 0 & 0
    \end{pmatrix} \ , \\
    \gamma^{(6d)}_1 &= \begin{pmatrix}
        0 & 0 & 0 & \sigma_1 \\
        0 & 0 & -\sigma_1 & 0 \\
        0 & - \sigma_1 & 0 & 0 \\
        \sigma_1 & 0 & 0 & 0
    \end{pmatrix} \ , \\
    \gamma^{(6d)}_2 &= \begin{pmatrix}
        0 & 0 & 0 & \sigma_2 \\
        0 & 0 & -\sigma_2 & 0 \\
        0 & - \sigma_2 & 0 & 0 \\
        \sigma_2 & 0 & 0 & 0
    \end{pmatrix} \ , \\
    \gamma^{(6d)}_3 &= \begin{pmatrix}
        0 & 0 & 0 & \sigma_3 \\
        0 & 0 & -\sigma_3 & 0 \\
        0 & - \sigma_3 & 0 & 0 \\
        \sigma_3 & 0 & 0 & 0
    \end{pmatrix} \ , \\
    \gamma_4^{(6d)} &= \begin{pmatrix}
        0 & 0 & \bbm{1}_2 & 0 \\
        0 & 0 & 0 & \bbm{1}_2 \\
        \bbm{1}_2 & 0 & 0 & 0 \\
        0 & \bbm{1}_2 & 0 & 0
    \end{pmatrix} \ , \\
    \gamma_5^{(6d)} &= \begin{pmatrix}
        0 & 0 & -i\bbm{1}_2 & 0 \\
        0 & 0 & 0 & i\bbm{1}_2 \\
        i\bbm{1}_2 & 0 & 0 & 0 \\
        0 & -i\bbm{1}_2 & 0 & 0
    \end{pmatrix} \ ,
\end{align}
\end{subequations} 
in 6d; it is simple to check that these both satisfy
\begin{equation}
    \{ \gamma_a , \gamma_b \} = 2 \eta_{ab} \ .
\end{equation}

\section{Derivation of the Lorentz Transformation Generator}
\label{sec: gen}
Let us derive the $\frak{so}(1,3)$ element corresponding to the transformation \eqref{eq: lambda}. We will do this using the accidental isomorphism
\begin{equation}
    SO(1,3) \cong \frac{SL(2,\bb{C})}{\bb{Z}_2} \ .
\end{equation}
of the 4d Lorentz group. Explicitly, given $\Lambda \in SO(1,3)$ we have two corresponding $SL(2,\bb{C})$ group elements
\begin{equation}
    S[\Lambda] = \pm \frac{\tensor{\Lambda}{^a_b} \sigma_a \Bar{\sigma}^b}{\sqrt{\det\brac{\tensor{\Lambda}{^a_b} \sigma_a \Bar{\sigma}^b}}} \ ,
\end{equation}
with
\begin{equation}
    \sigma_a = (\mathbbm{1}, \sigma_i) \; , \; \Bar{\sigma}_a = (-\mathbbm{1}, \sigma_i) \ .
\end{equation}
Since we require that $\Lambda = \bbm{1}$ maps to the identity on both sides, we must pick the positive sign. Using the components of $\Lambda$ given in \eqref{eq: lambda}, we find
\begin{equation} \label{eq: Weyl S}
    S[\Lambda] = \frac{e^{-\frac{ i x^+ }{4R}}}{4 \sqrt{\cos \brac{\frac{x^+}{2R}} }}
    \begin{pmatrix}
        3+e^{\frac{i x^+}{R}} - \frac{u}{\sqrt{2}R} &  -1+e^{\frac{i x^+}{R} } - \frac{u}{\sqrt{2}R} \\
        -1+e^{\frac{i x^+}{R} } + \frac{u}{\sqrt{2}R} & 3+e^{\frac{i x^+}{R}} + \frac{u}{\sqrt{2}R}
    \end{pmatrix} \ .
\end{equation}

The matrix $S[\Lambda]$ is the (positive-chirality) Weyl-spinor representation of $\Lambda$, so if $\Lambda$ is generated by the Lie algebra element $\lambda$, i.e.
\begin{equation}
    \Lambda = \Exp\brac{\lambda} \ ,
\end{equation}
then we have
\begin{equation}
    S[\Lambda] = \Exp\brac{\inv{4} \lambda^{ab} \sigma_{ab}} \ ,
\end{equation}
using the notation
\begin{equation}
    \sigma_{ab} = \inv{2} \brac{\sigma_a \bar{\sigma}_b - \sigma_b \bar{\sigma}_a} \ .
\end{equation}
It will be useful to rewrite the exponent in terms of three complex variables $\{\lambda_i\}$ by defining
\begin{align} \nonumber
    \lambda^{ab}\sigma_{ab} &= 2(\lambda^{01} + i \lambda^{23})\sigma_1 + 2(\lambda^{02} + i \lambda^{31}) \sigma_2 + 2 (\lambda^{03} + i \lambda^{12}) \sigma_3 \\ 
    &\equiv 2 \sum_i \lambda_i \sigma_i \ .
\end{align}
It is then easy to take the exponential of this, and we obtain
\begin{equation}
    S[\Lambda] = \begin{pmatrix}
        \cosh\brac{\frac{v}{2}} + \frac{\lambda_3 \sinh\brac{\frac{v}{2}}}{v} & \frac{(\lambda_1 - i \lambda_2) \sinh \brac{\frac{v}{2}}}{v} \\
        \frac{(\lambda_1 + i \lambda_2) \sinh \brac{\frac{v}{2}}}{v} & \cosh\brac{\frac{v}{2}} - \frac{\lambda_3 \sinh\brac{\frac{v}{2}}}{v}
    \end{pmatrix}
\end{equation}
with
\begin{equation}
    v = \sqrt{\lambda_1^2 + \lambda_2^2 + \lambda_3^2} \ .
\end{equation}
Comparing this with \eqref{eq: Weyl S} gives the three relations
\begin{subequations}
\begin{gather}
    \frac{\lambda_3  \sinh \brac{\frac{v}{2}}}{v} = - \frac{u\, e^{-\frac{i x^+}{4R}}}{4R \sqrt{2 \cos\brac{\frac{x^+}{2R}}}} \ , \\
    \frac{i \lambda_2  \sinh \brac{\frac{v}{2}}}{v} = \frac{u\, e^{-\frac{i x^+}{4R}}}{4R \sqrt{2 \cos\brac{\frac{x^+}{2R}}}} \ , \\
    \cosh\brac{\frac{v}{2}} = \frac{e^{-\frac{i x^+}{4R}} \brac{3 + e^{\frac{i x^+}{R}}} }{4\sqrt{\cos\brac{\frac{x^+}{2R}}}} \ .
\end{gather}
\end{subequations}
As the first two imply
\begin{equation}
    \lambda_3 = - i \lambda_2 \ ,
\end{equation}
we have $v = \lambda_1$. Its value is then fixed by considering the final relation; a quick calculation shows that
\begin{equation}
    \lambda_1 = \ln \cos\brac{\frac{x^+}{2R}} + \frac{i x^+}{2R} \ .
\end{equation}
Substituting this back into the other two relations and reorganising gives the Lie algebra element's components
\begin{subequations}
\begin{align}
    \lambda^{01} &= \ln\cos\brac{\frac{x^+}{2R}} \ , \\
    \lambda^{23} &= \frac{x^+}{2R} \ , \\ \nonumber
    \lambda^{03} = \lambda^{31} = \inv{2\sqrt{2}R \sin\brac{\frac{x^+}{2R}}} \Bigg[&x^2 \brac{\sin\brac{\frac{x^+}{2R}} \ln\cos\brac{\frac{x^+}{2R}} - \frac{x^+}{2R} \cos \brac{\frac{x^+}{2R}}} \\ & \hspace{-0.7cm} - x^3\brac{\cos\brac{\frac{x^+}{2R}} \ln\cos\brac{\frac{x^+}{2R}} + \frac{x^+}{2R} \sin \brac{\frac{x^+}{2R}}}\Bigg] \ , \\ \nonumber
    \lambda^{12} = - \lambda^{02} = \frac{1}{2\sqrt{2}R \sin\brac{\frac{x^+}{2R}}} \Bigg[&x^3 \brac{\sin\brac{\frac{x^+}{2R}} \ln\cos\brac{\frac{x^+}{2R}} - \frac{x^+}{2R}\cos\brac{\frac{x^+}{2R}}} \\  
    & \hspace{-0.7cm} + x^2\brac{\cos\brac{\frac{x^+}{2R}} \ln\cos\brac{\frac{x^+}{2R}} + \frac{x^+}{2R}\sin\brac{\frac{x^+}{2R}}}\Bigg] \ .
\end{align}
\end{subequations}
We can rewrite this covariantly as
\begin{subequations}
\begin{align}
    \lambda^{01} &= \ln\cos\brac{\frac{x^+}{2R}} \ , \\ \nonumber
    \lambda^{0i} = - \lambda^{1i} = - \inv{2\sqrt{2}R \sin\brac{\frac{x^+}{2R}}} \Bigg[&x^i \brac{\cos\brac{\frac{x^+}{2R}} \ln\cos\brac{\frac{x^+}{2R}} + \frac{x^+}{2R} \sin\brac{\frac{x^+}{2R}}} \\ 
    &\hspace{-1.5cm}+ R \Omega_{ij}x^j \brac{\sin\brac{\frac{x^+}{2R}} \ln\cos\brac{\frac{x^+}{2R}} - \frac{x^+}{2R}\cos\brac{\frac{x^+}{2R}}}\Bigg] \ , \\
    \lambda^{ij} &= \frac{\Omega_{ij} x^+}{2} \ .
\end{align}
\end{subequations}
A straightforward calculation shows that if we use extend this to 6d and use it as the generator for a 6d Lorentz group element we recover \eqref{eq: lambda}, so this is indeed the correct generator to consider.

\section{Normalisation of Scalar Two-Point Functions}
\label{sec: scalar coefficients}

\subsection{3d Scalar Fields} \label{sec: scalar coefficients 3d}
Let us consider the action for a mode of a free complex scalar field,
\begin{equation}
    S_k = 2\pi R \int d^3x \Bigg[ \frac{2ik}{R} \bar{\phi}_k \partial_- \phi_k + \bar{\phi}_k D^2 \phi_k\Bigg] \ .
\end{equation}
We will ultimately require $k\in\bb{Z} + \inv{2}$ but shall only constrain it to be real for now. The equation of motion of the mode is
\begin{equation}
    \cal{L}_k \phi_k \equiv \frac{2ik}{R} \partial_- \phi_k + D^2 \phi_k = 0 \ ,
\end{equation}
which has the obvious spherically-symmetric solution
\begin{equation}
    \phi_k = c_k \inv{\sqrt{z \bar{z}}} \brac{\frac{z}{\bar{z}}}^k
\end{equation}
away from the origin. The time-ordered two-point function will satisfy the equation of motion away from coincident insertion points, so we will also have
\begin{equation}
    G_k \equiv \vac{T\{\phi_k(x^-,x) \bar{\phi}_k(0)\}} = c_k \inv{\sqrt{z \bar{z}}} \brac{\frac{z}{\bar{z}}}^k
\end{equation}
up to a time-ordering factor that we will include below. We would like to fix the constant $c_k$. Standard QFT lore tells us that the time-ordered two-point function is the Green's function of $\cal{L}$, and we should have
\begin{equation} \label{eq: greens but wrong}
    \cal{L}_k G_k = \frac{i}{2\pi R} \delta(x^-) \delta^{(2)}(x) \ .
\end{equation}

Let us calculate the left hand side of the Green's function equation. This is most easily done by introducing the coordinates
\begin{equation}
    z = r e^{i\theta}
\end{equation}
with $r\in[0,\infty)$ and $\theta\in(0,\pi)$. With these, integration of spherically-symmetric functions over spacetime takes the form
\begin{equation}
    \int_{\bb{R}^3} d^3x = 4\pi R \int_{0}^{\infty} dr \int_0^{\pi} d\theta \, r \ ,
\end{equation}
and the two-point function is
\begin{equation} \label{eq: Gk}
    G_k = \frac{c_k F_k(\theta) e^{2i k \theta}}{r} \ .
\end{equation}
We have introduced the time-ordering factor $F_k(\theta)$ defined by
\begin{equation}
    F_k(\theta) = \begin{cases}
        1  &  \theta \in (\alpha,\beta) \\
        0  &  \text{otherwise}
    \end{cases} \ ,
\end{equation}
which restricts us to the interval with endpoints $\alpha$ and $\beta$. We will keep these arbitrary at this point.

The operator $\cal{L}_k$ is given in $(r,\theta)$ coordinates by
\begin{align} \nonumber
    \cal{L}_k =& - \frac{i}{R} \Bigg[\inv{2}\brac{e^{i\theta} - e^{-i\theta}} \brac{r \partial_r^2  + \partial_r  + \inv{r} \partial_{\theta}^2 } - e^{i\theta} \brac{k-\inv{2}} \brac{ \partial_r  + \frac{i}{r}\partial_{\theta} } \\ 
    &- e^{-i\theta} \brac{k+\inv{2}} \brac{\partial_r - \frac{i}{r} \partial_{\theta} } \Bigg] \ .
\end{align}

As $G_k$ is not differentiable at $r=0$ and $\theta=\alpha,\beta$, we need to introduce the notion of the distributional derivative\footnote{See \cite{Friedlander_Joshi_2003} for an introduction to working with differentiation of distributions.} to interpret \eqref{eq: greens but wrong}. Let $G$ and $f$ be some smooth functions on $\bb{R}^3$, and define the action of $G$ on $f$ by
\begin{equation}
    G[f] = \int_{\bb{R}^3} d^3x \, G f \ .
\end{equation}
As $G$ is smooth, the derivative $\cal{L}_kG$ is well-defined; we can then act with this on $f$ and integrate by parts to obtain
\begin{align} \nonumber
    \cal{L}_kG[f] = -2\pi i \int_{0}^{\infty} dr \int_0^{\pi} d\theta \, G \bigg[& (e^{i\theta} - e^{-i\theta}) \brac{r^2 \partial_r^2 f + 3r \partial_r f + \partial_{\theta}^2 f} + 2i (e^{i\theta} + e^{-i\theta} ) \partial_{\theta} f   \\ \nonumber
    & + 2i \brac{e^{i\theta} \brac{k-\inv{2}} - e^{-i\theta} \brac{k+\inv{2}}} \partial_{\theta} f \\ 
    & + 2 \brac{e^{i\theta}\brac{k-\inv{2}} + e^{-i\theta} \brac{k+\inv{2}}} r \partial_r f 
    \bigg] \ .
\end{align}
We see that this expression makes sense even when $G$ is non-differentiable; we then define the regularised distributional derivative of $G_k$ through the action on a test function $f$, with
\begin{equation}
    (\cal{L}_kG_k[f])_{\epsilon} = \int_{\epsilon}^{\infty} dr \int_0^{\pi} G_k \cal{L}_k^{\dag} f \ ,
\end{equation}
where $\cal{L}_k^{\dag}$ is defined by the previous expression.

We can now subsitute the expression \eqref{eq: Gk} for $G_k$ in this expression. As $G_k$ was chosen to satisfy $\cal{L}_kG_k = 0$ away from the singular points, the integral will only pick up boundary contributions. A brief computation gives
\begin{align} \nonumber
    (\cal{L}_kG_k[f])_{\epsilon} = -2\pi i c_k \int_{\epsilon}^{\infty} dr \int_{\alpha}^{\beta} d\theta \Bigg[& \frac{\partial}{\partial r} \bigg( \brac{ e^{i(2k+1) \theta} - e^{i (2k-1) \theta} }( r \partial_r f + 2 f) \\ \nonumber
    &+ 2 \brac{ e^{i(2k+1) \theta} \brac{k-\inv{2}} + e^{i (2k-1) \theta} \brac{k+\inv{2}} }f \bigg) \\ 
    & + \frac{\partial}{\partial \theta} \bigg( \brac{ e^{i(2k+1) \theta} - e^{i (2k-1) \theta} } \inv{r} \partial_{\theta} f \bigg)
    \Bigg] 
\end{align}
which evaluates to
\begin{align} \label{eq: useful anchor} \nonumber
    (\cal{L}_kG_k[f])_{\epsilon} = &2\pi c_k \delta_0[f] \bigg[ e^{i(2k+1)\theta} + e^{i(2k-1)\theta} \bigg]_{\theta = \alpha}^{\theta = \beta} \\ 
    &+ 4\pi c_k \int_{\epsilon}^{\infty} \frac{dr}{r} \bigg[ e^{2ik\theta} \sin\theta \partial_{\theta} f \bigg]_{\theta = \alpha}^{\theta = \beta}
    + O(\epsilon) \ .
\end{align}

We see that for generic values of $\alpha$ and $\beta$ we get a non-delta function term that survives as we take $\epsilon\to 0$. The most obvious solution to this is to take $\alpha = 0$ and $\beta = \pi$: this mirrors the time-ordered two-point function of a Lorentzian CFT, where no $\Theta$-functions are present. With this choice, we can take $\epsilon\to 0$ to find
\begin{equation} \label{eq: DG a=0 b = pi}
    \cal{L}_kG_k = -4\pi c_k  \brac{1 + e^{2\pi i k}} \delta(x^-) \delta^{(2)}(x) \ .
\end{equation}
For generic $k$ we have no issues; comparing with \eqref{eq: greens but wrong} then gives
\begin{equation}
    c_k = - \frac{i}{8\pi^2 R \brac{1 + e^{2\pi i k}}}
\end{equation}
as the coefficient of the two-point function. However, the right hand side of \eqref{eq: DG a=0 b = pi} vanishes when $k\in\bb{Z}+\inv{2}$, which are precisely the values we are interested in! 

To find the resolution of this, we need to think about the 4d origin of the theory; since $\ket{\Omega}$ is the vacuum of a free Lorentzian scalar, it must be annihilated by all lowering operators. In particular, this applies to the modes arising from the reduction on $x^+$, as these are eigenstates of $p_+$. As discussed in \cite{Lambert:2021nol}, there is a slight subtlety here: the split into raising and lowering operators on $\ket{\Omega}$ is naturally defined for the modes of $\hat{\phi}$, which for the 3d theory are related to the modes of $\phi$ by
\begin{equation}
    \hat{\phi}_{k + \inv{2}} = \inv{2} \brac{\phi_{k} + \phi_{k+1}} \ .
\end{equation}
Requiring our theory makes sense as the reduction of a well-defined 4d theory imposes
\begin{equation} \label{eq: DLCQ conditions}
    \hat{\phi}_l \ket{\Omega} = \bar{\hat{\phi}}_{-l} \ket{\Omega} = 0 \ \forall \, l>0 \ ,
\end{equation}
which is equivalent to the conditions
\begin{equation}
    \phi_k \ket{\Omega} = \bar{\phi}_{-k}\ket{\Omega} = 0 \ \forall \, k>-\inv{2} \ .
\end{equation}

Taking time-ordering with respect to $x^-$, we then have
\begin{equation}
    \vac{T\{ \phi_k(x^-,x) \bar{\phi}_k(0) \} } = \begin{cases}
        \Theta(x^-) \vac{\phi_k(x^-,x) \bar{\phi}_k(0)} &  k > 0 \\
        \Theta(-x^-) \vac{ \bar{\phi}_k(0) \phi_k(x^-,x) } & k < 0
    \end{cases} \ .
\end{equation}
The two-point functions of the modes are related to the two-point function of the 4d field by
\begin{equation} \label{eq: greens exp}
    \vac{T\{ \hat{\phi}(\hat{x}) \hat{\bar{\phi}}(0) \} } = \cos \brac{\frac{x^+}{2R}} \sum_{k\in\bb{Z}+\inv{2}} e^{- \frac{ik x^+}{R}} G_k(x^-,x) \ .
\end{equation}
The 4d field satisfies the Green's function equation
\begin{equation} \label{eq: 4d green}
    \hat{\square} \vac{T\{ \hat{\phi}(\hat{x}) \hat{\bar{\phi}}(0) \} } = i \delta^{(4)}(\hat{x}) \ ,
\end{equation}
where the d'Alembertian $\hat{\square} = \hpart^2$ is given by
\begin{equation}
    \hat{\square} = \cos^2\brac{\frac{x^+}{2R}} \brac{-2 \brac{\partial_+ + \frac{\tan\brac{\frac{x^+}{2R}}}{2R}} \partial_- + D^2}
\end{equation}
after transforming to un-hatted coordinates. The factor of $\cos^2\brac{\frac{x^+}{2R}}$ is evaluated on a delta function, and is therefore unity. Substituting the expansion \eqref{eq: greens exp} into the equation \eqref{eq: 4d green} and using the periodic delta function
\begin{equation}
    \delta(x^+) = \inv{2\pi R} \sum_{l\in\bb{Z}} e^{-\frac{i l x^+}{R}}
\end{equation}
to reduce on $x^+$, we find the infinite tower of relations
\begin{equation} \label{eq: greens but right}
    \cal{L}_{l - \inv{2}} G_{l - \inv{2}} + \cal{L}_{l + \inv{2}} G_{l + \inv{2}} = \frac{i}{\pi R} \delta(x^-) \delta^{(2)}(x)
\end{equation}
for $l\in\bb{Z}$.

We see that the set of equations \ref{eq: greens but right} impose weaker constraints than the set \ref{eq: greens but wrong}. However, they share one important property. If we examine the equation for $l=0$ away from $x^-=0$, we have
\begin{equation}
    \cal{L}_{-\inv{2}} G_{-\inv{2}} + \cal{L}_{\inv{2}} G_{\inv{2}} = 0 \ .
\end{equation}
However, $G_{\inv{2}}$ and $G_{-\inv{2}}$ are only non-trivial for non-overlapping ranges of $x^-$. This means that if we restrict our attention to $x^->0$ we find the equation
\begin{equation}
    \cal{L}_{\inv{2}} G_{\inv{2}} = 0 \ ,
\end{equation}
whereas if we focus on $x^-<0$ we find the analogous statement for $G_{-\inv{2}}$. This propagates through the tower of equations, and we see that all the two-point functions satisfy the classical equations of motion away from singularities. 

We will now use the modified set of equation to fix $c_k$. It is necessary to consider the cases $l>0$, $l=0$, and $l<0$ separately. Let us start with $l>0$, which encompasses the two-point functions with $k>0$. In this regime, both correlators will be of the form
\begin{equation}
    G_k = \begin{cases}
        \frac{c_k e^{2ik\theta}}{r} &  \theta \in \brac{0, \frac{\pi}{2}} \\
        0 & \text{otherwise}
    \end{cases} \ .
\end{equation}
Using \eqref{eq: useful anchor}, we have
\begin{equation}
    (\cal{L}_k G_k[f])_{\epsilon} =  -4 c_k \pi \brac{ \delta_0[f] - \int_{\epsilon}^{\infty} \frac{dr}{r} \, e^{ik\pi} \brac{\partial_{\theta} f}_{\theta = \frac{\pi}{2}}} + O(\epsilon) \ .
\end{equation}
Since this means the left hand side of \eqref{eq: greens but right} becomes (after taking $\epsilon\to 0$)
\begin{equation}
    \cal{L}_{l-\inv{2}} G_{l-\inv{2}}[f] + \cal{L}_{l+\inv{2}} G_{l+\inv{2}}[f] = - 4\pi (c_{l-\inv{2}} + c_{l+\inv{2}}) \delta_0[f] - 4\pi i (c_{l-\inv{2}} - c_{l+\inv{2}}) \int_0^{\infty} \frac{dr}{r} e^{il \pi} \brac{\partial_{\theta} f}_{\theta = \frac{\pi}{2}} \ ,
\end{equation}
we see that we must take
\begin{equation} \label{eq: 4d coefficient k>0}
    c_k = - \frac{i}{8\pi^2 R}
\end{equation}
for $k>0$. The calculation for $l<0$ is almost identical; as we have
\begin{equation}
    G_k = \begin{cases}
        \frac{c_k e^{2ik\theta}}{r} & \theta \in \brac{\frac{\pi}{2},\pi} \\
        0 & \text{otherwise}
    \end{cases} 
\end{equation}
for $k<0$, \eqref{eq: useful anchor} evaluates to
\begin{equation}
    (\cal{L}_k G_k[f])_{\epsilon} = 4\pi c_k \brac{\delta_0[f] - \int_{\epsilon}^{\infty} \frac{dr}{r}\, e^{ik\pi} \brac{\partial_{\theta} f}_{\theta = \frac{\pi}{2}}} + O(\epsilon) \ .
\end{equation}
After taking $\epsilon\to0$ we see that \eqref{eq: greens but right} gives
\begin{equation}
    \cal{L}_{l-\inv{2}} G_{l-\inv{2}}[f] + \cal{L}_{l+\inv{2}} G_{l+\inv{2}}[f] =  4\pi (c_{l-\inv{2}} + c_{l+\inv{2}}) \delta_0[f] + 4\pi i (c_{l-\inv{2}} - c_{l+\inv{2}}) \int_0^{\infty} \frac{dr}{r} e^{il\pi} \brac{\partial_{\theta} f}_{\theta = \frac{\pi}{2}} \ ,
\end{equation}
so the conditions are satisfied if
\begin{equation} \label{eq: 4d coefficient k<0}
    c_k = \frac{i}{8\pi^2 R}
\end{equation}
for $k<0$. Finally, we must check the case $l=0$, which links the two previous results. Using our prior calculations we have
\begin{equation}
    \cal{L}_{-\inv{2}} G_{-\inv{2}}[f] + \cal{L}_{\inv{2}} G_{\inv{2}}[f] = 4\pi\brac{c_{-\inv{2}} - c_{\inv{2}}} \delta_0[f] + 4\pi i \brac{ c_{-\inv{2}} + c_{\inv{2}} } \int_{0}^{\infty} \frac{dr}{r}\, \brac{\partial_{\theta} f}_{\theta = \frac{\pi}{2}} \ .
\end{equation}
Substituting \eqref{eq: 4d coefficient k>0} and \eqref{eq: 4d coefficient k<0} into this gives
\begin{equation}
    \cal{L}_{-\inv{2}} G_{-\inv{2}}[f] + \cal{L}_{\inv{2}} G_{\inv{2}}[f] = \frac{i}{\pi R} \delta_0[f] \ ,
\end{equation}
which is consistent with \eqref{eq: greens but right}.

As a test of these results, we can resum \eqref{eq: greens exp} and check that we get the correct result. Using the $i\epsilon$ prescription described in section \ref{sec: corr} we find
\begin{align} \nonumber
    \vac{ T\{ \hat{\phi}(\hat{x}) \hat{\bar{\phi}}(0) \} } &= -\frac{i \cos\brac{\frac{x^+}{2R}}}{8\pi^2 R} \sum_{k=0}^{\infty} \brac{ \Theta(x^-) \brac{\frac{z e^{- \frac{i x^+}{R}}}{\bar{z}}}^{k+ \inv{2}} - \Theta(-x^-) \brac{\frac{z e^{- \frac{i x^+}{R}}}{\bar{z}}}^{-\brac{k+ \inv{2}}} } \\ \nonumber
    &= \frac{i \cos \brac{\frac{x^+}{2R}}}{8\pi^2 R} \inv{z e^{-\frac{i x^+}{2R}} - \bar{z} e^{\frac{i x^+}{2R}}} \\
    &= \inv{ 4\pi^2 \brac{- 2 \hat{x}^+ \hat{x}^- + \hat{x}^2}} \ , 
\end{align}
which is the two-point function of a canonically normalised 4d massless scalar field, as hoped.

\subsection{5d Scalar Fields} \label{sec: scalar coefficients 5d}
We can perform the exact same analysis for the reduction of a 6d scalar field to a 5d theory. We will use the same spherically-symmetric coordinates as in the 3d case. The two-point function obtained from the classical equations of motion is
\begin{equation}
    G_k = \frac{c_k F_k(\theta) e^{2ik\theta}}{r^2} \ ,
\end{equation}
where we will be interested in taking $k\in\bb{Z}$. Integrals over spherically-symmetric functions take the form
\begin{equation}
    \int_{\bb{R}^5} d^5x = \brac{4\pi R}^2 \int_0^{\infty} \int_{0}^{\pi} dr d\theta \, r^2 \sin\theta \ .
\end{equation}
In these coordinates, the classical equation of motion is
\begin{align} \nonumber
    \cal{L}_k G_k = \frac{-2i}{R} \Bigg[& \inv{4}\brac{e^{i\theta} - e^{-i\theta}} \brac{r \partial_r^2 G + \partial_r G + \inv{r} \partial_{\theta}^2 G} - \inv{2} e^{i\theta} (k-1) \brac{ \partial_r G + \frac{i}{r}\partial_{\theta} G}\\ 
    &- \inv{2} e^{-i\theta} (k+1) \brac{\partial_r G - \frac{i}{r} \partial_{\theta} G} \Bigg] \ .
\end{align}
The same procedure as above allows us to define the regularised distributional derivative of $G_{k}$ as 
\begin{align} \nonumber
    (\cal{L}_k G_k[f])_{\epsilon} = -4\pi^2 R \int_{\epsilon}^{\infty} dr \int_{0}^{\pi} d\theta \, G_k \Bigg[& \brac{e^{i\theta} - e^{-i\theta}}^2 \brac{r^3 \partial_r^2 f + 5 r^2 \partial_r f + 4rf + r \partial_{\theta}^2 f} \\ \nonumber
    & + 2 (r^2 \partial_r f + 2rf) \brac{(k-1) (e^{2i\theta} -1) + (k+1) (1- e^{-2i\theta})} \\ \nonumber
    &+ 4i \brac{e^{2i\theta} - e^{-2i\theta} } r\partial_{\theta} f - 4 \brac{e^{2i\theta} + e^{-2i\theta} } rf \\ \nonumber
    &+ 2ir \partial_{\theta} f \Big((k-1) (e^{2i\theta} -1) - (k+1) (1- e^{-2i\theta})\Big) \\
    &- 4rf \brac{(k-1) e^{2i\theta} - (k+1)  e^{-2i\theta}}
    \Bigg] \ .
\end{align}
Subsituting in our expression for $G_k$ gives
\begin{align} \nonumber
    \brac{\cal{L}_k G_k[f]}_{\epsilon} = - 4\pi^2 R \,c_k \int_{\epsilon}^{\infty} dr \int_{\alpha}^{\beta} d\theta \Bigg[& \frac{\partial}{\partial r} \bigg( e^{2i(k+1) \theta} \brac{r \partial_r f + 2(k+1)f} + e^{2i(k-1) \theta} \brac{r\partial_r f -2(k-1) f} \\ \nonumber
    &- 2 e^{2i k \theta} \brac{r \partial_r f + 2 f} \bigg) + \frac{\partial}{\partial \theta} \bigg( \inv{r} \partial_{\theta} f \bigg(e^{2i(k+1)\theta}  \\ 
    &+ e^{2i(k-1) \theta} - 2 e^{2ik \theta} \bigg) \bigg)
    \Bigg] \ ,
\end{align}
which localises to boundary contributions as expected. When none of the exponents vanish (i.e. $k\neq-1,0,1$) this is
\begin{equation}
    \brac{\cal{L}_k G_k[f]}_{\epsilon} = 8 \pi^2 R\, c_k \Bigg[ e^{2ik \theta} \bigg( \delta_0[f] \brac{\sin2\theta + \frac{i}{k}} - \brac{\cos2\theta - 1} \int_{\epsilon}^{\infty} \frac{dr}{r} \, \partial_{\theta} f \bigg) \Bigg]_{\theta = \alpha}^{\theta=\beta} + O(\epsilon) \ ,
\end{equation}
with the exceptional cases
\begin{subequations}
\begin{align}
    (\cal{L}_1 G_1)_{\epsilon} &= 4\pi^2 R \,c_1  \Bigg[ e^{2i\theta} \bigg( i  \delta_0[f]\brac{ 2 - e^{2i\theta} }  - 2   \brac{\cos2\theta - 1} \int_{\epsilon}^{\infty} \frac{dr}{r}\, \partial_{\theta} f \bigg) \Bigg]_{\theta = \alpha}^{\theta = \beta} + O(\epsilon)  \ , \\
    (\cal{L}_0 G_0)_{\epsilon} &= 8\pi^2 R \,c_0 \Bigg[ \delta_0[f] \brac{\sin2\theta - 2 \theta} - \brac{\cos2\theta - 1} \int_{\epsilon}^{\infty} \frac{dr}{r} \, \partial_{\theta} f  \Bigg]_{\theta = \alpha}^{\theta = \beta} + O(\epsilon) \ , \\
    (\cal{L}_{-1} G_{-1}[f])_{\epsilon} &= 4\pi^2 R \,c_{-1} \Bigg[ e^{-2i\theta} \bigg( i \delta_0[f] \brac{e^{-2i\theta} - 2}  - 2 \brac{\cos2\theta - 1} \int_{\epsilon}^{\infty} \frac{dr}{r}\, \partial_{\theta} f \bigg) \Bigg]_{\theta = \alpha}^{\theta = \beta} + O(\epsilon) \ .
\end{align}
\end{subequations}

For generic values of $k$, we can take $\alpha = 0$ and $\beta = \pi$ to obtain (after taking the $\epsilon\to 0$ limit)
\begin{equation}
    \cal{L}_k G_k[f] = \frac{8\pi^2 R \brac{e^{2\pi i k} - 1}}{k} c_k \delta_0[f] \ ,
\end{equation}
so the Green's function equation \eqref{eq: greens but wrong} is satisfied if we take
\begin{equation}
    c_k = \frac{k}{16\pi^3 R^2 \brac{e^{2\pi i k} - 1}} \ .
\end{equation}
However, like the 3d case this is not possible when $k\in\bb{Z}$, which is the case of interest.

Let us follow the same steps as before to obtain the correct equations for the tower of 5d two-point functions from the 6d Green's function equation. The time-ordered 6d two-point function is given in terms of the modes by
\begin{equation}
    \vac{T\{ \hat{\phi}(\hat{x}^+, \hat{x}^-, \hat{x}) \bar{\hat{\phi}}(0) \}} = \cos^2\brac{\frac{x^+}{2R}} \sum_{k\in \bb{Z}} e^{- \frac{i kx^+}{R} } G_k(x^-,x) \ .
\end{equation}
The modes of $\hat{\phi}$ are related to those of $\phi$ by
\begin{equation}
    \hat{\phi}_k = \inv{4} \brac{\phi_{k-1} + 2 \phi_k + \phi_{k+1}} \ ,
\end{equation}
so the conditions \eqref{eq: DLCQ conditions} become
\begin{equation}
    \phi_{k} \ket{\Omega} = \bar{\phi}_{-k} \ket{\Omega} = 0 \ \forall \, k>-1 \ .
\end{equation}
This means that the mode two-point functions are of the form
\begin{equation}
    G_k(x^-.x) = \begin{cases}
        \Theta(x^-) \vac{\phi_k(x^-,x) \bar{\phi}_k(0)}  & k>0 \\
        0  & k = 0 \\
        \Theta(-x^-) \vac{\bar{\phi}_k(0) \phi_k(x^-,x)}  & k<0
    \end{cases} \ .
\end{equation}
In particular, we immediately have the condition $c_0$ = 0.

Acting with the 6d d'Alembertian
\begin{equation}
    \hat{\square} = \cos^2\brac{\frac{x^+}{2R}} \brac{ - 2 \brac{\partial_+ + \inv{R} \tan\brac{\frac{x^+}{2R}}} + D^2 }
\end{equation}
and rearranging, we see that the Green's function equation
\begin{equation}
    \hat{\square} \vac{T\{ \hat{\phi}(\hat{x}^+, \hat{x}^-, \hat{x}) \bar{\hat{\phi}}(0) \}} = i \delta^{(6)}(\hat{x})
\end{equation}
reduces to the tower of equations
\begin{equation} \label{eq: 6d equation tower}
    \cal{L}_{k-1} G_{k-1} + \cal{L}_{k+1} G_{k+1} + 2 \cal{L}_k G_k = \frac{2i}{\pi R} \delta(x^-) \delta^{(4)}(x)
\end{equation}
for $k\in\bb{Z}$. Substituting in $\alpha = 0$, $\beta = \frac{\pi}{2}$ for $k>0$ and $\alpha = \pi$, $\beta = \frac{\pi}{2}$ for $k<0$, the distributional derivatives of the two-point functions are (taking $\epsilon \to 0$)
\begin{subequations}
\begin{equation}
    \cal{L}_k G_k[f] = \frac{8\pi^2 i R c_k}{k}\brac{(-1)^k - 1} \delta_0[f] + 16\pi^2 R (-1)^k c_k \int_0^{\infty} \frac{dr}{r} \brac{\partial_{\theta} f}_{\theta = \frac{\pi}{2}}
\end{equation}
for $k>0$, and
\begin{equation}
    \cal{L}_{k} G_{k}[f] = -\frac{8\pi^2 i R c_{k}}{k}\brac{(-1)^k - 1} \delta_0[f] - 16\pi^2 R (-1)^k c_k \int_0^{\infty} \frac{dr}{r} \brac{\partial_{\theta} f}_{\theta = \frac{\pi}{2}}
\end{equation}
\end{subequations}
for $k<0$. We see that the equations \eqref{eq: 6d equation tower} are satisfied if we take
\begin{equation} \label{eq: n=3 coeff}
    c_k = \begin{cases}
        - \frac{k}{16\pi^3 R^2} & k>0 \\
        0 & k=0 \\
        \frac{k}{16\pi^3 R^2} & k< 0
    \end{cases} \ .
\end{equation}
A similar calculation to the resummed 4d two-point function then gives\footnote{Suppressing the terms arising from the $i\epsilon$ prescription.}
\begin{equation}
    \vac{T\{\hat{\phi}(\hat{x}^+, \hat{x}^-, \hat{x}) \bar{\hat{\phi}}(0) \}} = \inv{4\pi^3 \brac{- 2 \hat{x}^+ \hat{x}^- + \hat{x}^i \hat{x}^i}^2 } \ , 
\end{equation}
which we recognise as the two-point function of a canonically normalised 6d scalar field.

\subsection{The DLCQ Limit of \texorpdfstring{$SU(1,n)$}{SU(1,n)} Scalar Fields}
\label{sec: appendix DLCQ}

Let us see how we recover the familiar DLCQ picture from this. As discussed in section \ref{sec: DLCQ}, we can think of the DLCQ limit by writing $k$ and $R$ as
\begin{subequations}
\begin{align}
    k &=  N k_+ \ , \\
    R &= N R_+ \ , 
\end{align}
\end{subequations}
and taking the limit $N\to\infty$ keeping $k_+$ and $R_+$ fixed. To get a finite action after this limit we must also rescale the mode fields by $\inv{\sqrt{N}}$. Since we have
\begin{equation}
    \frac{k + c}{R} = \frac{k_+}{R_+} + O\Big(\inv{N}\Big)
\end{equation}
for any $c$ that doesn't scale with $N$, sums over $k\in\bb{Z}+\alpha$ become sums over $k_+\in\bb{Z}$ in the $N\to\infty$ limit. From here onwards we will drop pluses for ease of notation. The action for the DLCQ reduction of a complex scalar field in $2n$-dimensions can then be seen to be
\begin{equation} \label{eq: DLCQ action}
    S = 2\pi R \sum_{k\in\bb{Z}} \int dx^- d^{2n-2}x \brac{ \frac{2ik}{R}\bar{\phi}_k \partial_- \phi_k + \Bar{\phi}_k \partial^2 \phi_k } \ .
\end{equation}
We observe that in the DLCQ limit the Green's function-like equations \eqref{eq: greens but right} and \eqref{eq: 6d equation tower} both reduce to the standard Schr\"odinger field Green's function equation
\begin{equation} \label{eq: DLCQ greens}
    \brac{\frac{2ik}{R} \partial_- + \partial^2} \vac{T\{ \phi_k(x^-,x) \bar{\phi}_k(0)\}} = \frac{i}{2\pi R} \delta(x^-) \delta^{2n-2}(x) \ ,
\end{equation}
which is the obvious set of equations obtained from the action \eqref{eq: DLCQ action}. The solution to this is well known, and we briefly review it. Using the fact that the two-point function satisfies the classical equation of motion away from the origin and the DLCQ vacuum condition
\begin{equation}
    \phi_k \ket{\Omega} = \bar{\phi}_{-k} \ket{\Omega} = 0
\end{equation}
for $k>0$, we find
\begin{equation} \label{eq: dlcq 2 point 1}
    G_k \equiv \vac{T\{ \phi_k(x^-,x) \bar{\phi}_k(0)\}} = \begin{cases}
        c_k \Theta(x^-)(x^-)^{1-n} \exp\brac{\frac{ik x^2}{2R x^-}}   & k>0 \\
        0  & k = 0 \\
        c_k \Theta(-x^-)(x^-)^{1-n} \exp\brac{\frac{ik x^2}{2R x^-}}  & k < 0
    \end{cases} \ .
\end{equation}
Let us focus on the $k>0$ sector. Since $G_k$ is only non-differentiable at $x^- = 0$, we define the distributional derivative as
\begin{equation}
    (\cal{L}_k G_k)_{\epsilon} = \brac{\int_{-\infty}^{-\epsilon} dx^- + \int_{\epsilon}^{\infty} dx^- } \int_{\bb{R}^{2n-2}} d^{2n-2}x \brac{ - \frac{2ik}{R} \partial_- f + \partial^2 f} G_k 
\end{equation}
where we use $\cal{L}_k$ to denote the differential operator in \eqref{eq: DLCQ greens}. Substituting in our expression for $G_k$ and formally treating our spatial integrals as if they are convergent we find
\begin{align} \nonumber
    (\cal{L}_k G_k)_{\epsilon} &= \frac{2ik }{R } \epsilon^{1-n} c_k \, \delta_0[f] \brac{ \int_{\bb{R}} dx\, e^{\frac{ik x^2}{2R \epsilon}}}^{2n-2} + O(\epsilon) \\
    &= \frac{2ik}{R} \brac{-\frac{2\pi R}{i k}}^{n-1} c_k \, \delta_0[f] + O(\epsilon) \ .
\end{align}
We can now take $\epsilon\to0$ and compare with \eqref{eq: DLCQ greens} to find the coefficient
\begin{equation} \label{eq: DLCQ coeff 1}
    c_k = \inv{4\pi k} \brac{-\frac{i k}{2\pi R}}^{n-1}
\end{equation}
for $k>0$. Similarly, working through the same calculation for the $k<0$ sector gives
\begin{equation} \label{eq: DLCQ coeff 2}
    c_k = - \inv{4\pi k} \brac{-\frac{ik}{2\pi R}}^{n-1} \ .
\end{equation}
We now consider directly taking the DLCQ limit of the $SU(1,n)$ two-point functions. We saw above that for a complex scalar field these are of the form
\begin{equation}
    G_k = \frac{c_k F_k(\theta) e^{2ik \theta}}{r^{n-1}} \ ,
\end{equation}
where
\begin{subequations}
\begin{align}
    r &= \sqrt{(x^-)^2 + \brac{\frac{x^2}{4R}}^2} \ , \\
    \theta &= - \frac{i}{2} \ln \brac{\frac{1 + \frac{i x^2}{4 R x^-}}{1 - \frac{i x^2}{4R x^-}}} \ .
\end{align}
\end{subequations}
In the DLCQ limit we have
\begin{subequations}
\begin{align}
    r &= x^- + O\brac{\inv{R}} \ , \\
    \theta &= \frac{x^2}{4R x^-} + O\brac{\inv{R^2}} \ ,
\end{align}
\end{subequations}
so the limit of the two-point function is (assuming that no limit needs to be taken in the coefficients)
\begin{equation}
    G_k \to \begin{cases}
        c_k \Theta(x^-) (x^-)^{1-n} \exp\brac{\frac{ i k x^2}{2R x^-}} & k > 0 \\
        0 & k= 0 \\
        c_k \Theta(-x^-) (x^-)^{1-n} \exp\brac{\frac{ i k x^2}{2R x^-}} & k < 0
    \end{cases} \ ,
\end{equation}
where we add the $k=0$ case in by hand for $n=2$. As hoped, this is identical to the DLCQ correlation functions in \eqref{eq: dlcq 2 point 1} as long as the coefficients agree in the $SU(1,n)$ and DLCQ calculations. Comparing \eqref{eq: DLCQ coeff 1} and \eqref{eq: DLCQ coeff 2} with \eqref{eq: 4d coefficient k>0} and \eqref{eq: 4d coefficient k<0} for $n=2$ and \eqref{eq: n=3 coeff} for $n=3$, we see that this is indeed true.

It will be interesting to resum the modes and obtain the higher-dimensional DLCQ two-point function. Doing this gives
\begin{align} \nonumber
    \vac{T \{ \phi(x^+,x^-,x) \bar{\phi}(0)\}} &= \sum_{k = 1}^{\infty} \brac{e^{- \frac{i k x^+}{R}} G_k + e^{\frac{i k x^+}{R}} G_{-k} } \\
    &= \inv{4\pi} \brac{ \frac{-i}{2\pi R x^-}}^{n-1} \brac{\Theta(x^-) \polylog{2-n}{q} + (-1)^{n-1} \Theta(-x^-) \polylog{2-n}{q^{-1}}} \ ,
\end{align}
where we have defined $q$ as in \eqref{eq: q def} and used the usual $i\epsilon$ prescription \eqref{eq: i epsilon} to ensure convergence of the sum. We can then use the identity\footnote{Note that this means we are restricting our attention to the DLCQ of scalar field theories in even spacetime dimensions.} \eqref{eq: polylog identity} to simplify the two-point function to
\begin{equation}
    \vac{T \{ \phi(x^+,x^-,x) \bar{\phi}(0)\}} = \inv{4\pi} \brac{\frac{-i}{2\pi R x^-}}^{n-1} \polylog{2-n}{q} \ .
\end{equation}
For the exceptional $n=2$ case we instead have
\begin{equation}
    \vac{T \{ \phi(x^+,x^-,x) \bar{\phi}(0)\}} =  
    \inv{16\pi^2 R} \brac{\inv{x^-}\cot\brac{\frac{x_{\mu} x^{\mu}}{4R \, x^-}} + \frac{i}{\abs{x^-}}} \ ,
\end{equation}
where as before
\begin{equation}
    x_{\mu} x^{\mu} \equiv -2x^+ x^- + x^i x^i \ .
\end{equation}
In order to recover the Minkowski two-point function we need to take the $R\to\infty$ limit to decompactify the null direction; using the large-$R$ expansions \eqref{eq: large R expansions} we see that in both cases we get
\begin{equation}
    \lim_{R\to\infty} \Big(\vac{T \{ \phi(x^+,x^-,x) \bar{\phi}(0)\}}\Big) = \frac{(n-2)!}{4\pi^n \brac{x_{\mu} x^{\mu}}^{n-1}} \ ,
\end{equation}
which we recognise as the canonically-normalised two-point function of a massless scalar field in $2n$-dimensional Minkowski spacetime.

\bibliography{references.bib}

\end{document}